\documentclass[useAMS,usenatbib,usegraphicx]{mn2e}
\usepackage{deluxetable,lscape}

\title[A stellar autopsy of the radio-bright \hbox{SN\,1996cr}]{Performing a stellar autopsy using the radio-bright remnant of \hbox{SN\,1996cr}
} 
\author[C. Meunier et al.]{\parbox{\textwidth}{C. Meunier$^{1}$\thanks{E-mail: clmeunie@uc.cl}, 
F. E. Bauer$^{1,2,3}$\thanks{fbauer@astro.puc.cl}, 
V. V. Dwarkadas$^{4}$\thanks{vikram@oddjob.uchicago.edu}, 
B. Koribalski$^{5}$,
B. Emonts$^{5}$, 
R. W. Hunstead$^{6}$,
D. Campbell-Wilson$^{6}$
C. Stockdale$^{7}$,
S. J. Tingay$^{8}$}\vspace{0.4cm}\\
$^{1}$Pontificia Universidad Cat\'{o}lica de Chile, Departamento de Astronom\'{\i}a y Astrof\'{\i}sica, Casilla 306, Santiago 22, Chile\\
$^{2}$Space Science Institute, 4750 Walnut Street, Suite 205, Boulder, Colorado 80301\\
$^{3}$Columbia Astrophysics Laboratory, 550 W. 120th St., Columbia University, New York, NY 10027\\
$^{4}$Department  of Astronomy and Astrophysics, U Chicago, 5640 S Ellis Ave, Chicago, IL 60637\\
$^{5}$Australia Telescope National Facility, CSIRO Astronomy and Space Science, PO Box 76, Epping, NSW, 1710, Australia\\
$^{6}$Sydney Institute for Astronomy, School of Physics, University of Sydney, NSW 2006, Australia\\
$^{7}$Department of Physics, Marquette University, P.O. Box 1881, Milwaukee, WI 53201-1881, USA\\
$^{8}$International Centre for Radio Astronomy Research, Curtin University, Bentley 6102, WA, Australia}
\begin{document}
\newcommand{\vper}{\mbox{${v_{\perp}}$}}
\newcommand{\vpar}{\mbox{${v_{\parallel}}$}}
\newcommand{\uper}{\mbox{${u_{\perp}}$}}
\newcommand{\vperout}{\mbox{${{v_{\perp}}_{o}}$}}
\newcommand{\uperout}{\mbox{${{u_{\perp}}_{o}}$}}
\newcommand{\vperin}{\mbox{${{v_{\perp}}_{i}}$}}
\newcommand{\uperin}{\mbox{${{u_{\perp}}_{i}}$}}
\newcommand{\upar}{\mbox{${u_{\parallel}}$}}
\newcommand{\uparout}{\mbox{${{u_{\parallel}}_{o}}$}}
\newcommand{\vparout}{\mbox{${{v_{\parallel}}_{o}}$}}
\newcommand{\uparin}{\mbox{${{u_{\parallel}}_{i}}$}}
\newcommand{\vparin}{\mbox{${{v_{\parallel}}_{i}}$}}
\newcommand{\dout}{\mbox{${\rho}_{o}$}}
\newcommand{\din}{\mbox{${\rho}_{i}$}}
\newcommand{\da}{\mbox{${\rho}_{1}$}}
\newcommand{\mfast}{\mbox{$\dot{M}_{f}$}}
\newcommand{\mslow}{\mbox{$\dot{M}_{a}$}}
\newcommand{\beqn}{\begin{eqnarray}}
\newcommand{\eeqn}{\end{eqnarray}}
\newcommand{\be}{\begin{equation}}
\newcommand{\ee}{\end{equation}}
\newcommand{\noi}{\noindent}
\newcommand{\ftheta}{\mbox{$f(\theta)$}}
\newcommand{\gtheta}{\mbox{$g(\theta)$}}
\newcommand{\ltheta}{\mbox{$L(\theta)$}}
\newcommand{\stheta}{\mbox{$S(\theta)$}}
\newcommand{\utheta}{\mbox{$U(\theta)$}}
\newcommand{\xitheta}{\mbox{$\xi(\theta)$}}
\newcommand{\vs}{\mbox{${v_{s}}$}}
\newcommand{\ro}{\mbox{${R_{0}}$}}
\newcommand{\pa}{\mbox{${P_{1}}$}}
\newcommand{\va}{\mbox{${v_{a}}$}}
\newcommand{\vo}{\mbox{${v_{o}}$}}
\newcommand{\vp}{\mbox{${v_{p}}$}}
\newcommand{\vw}{\mbox{${v_{w}}$}}
\newcommand{\vf}{\mbox{${v_{f}}$}}
\newcommand{\lprime}{\mbox{${L^{\prime}}$}}
\newcommand{\uprime}{\mbox{${U^{\prime}}$}}
\newcommand{\sprime}{\mbox{${S^{\prime}}$}}
\newcommand{\xiprime}{\mbox{${{\xi}^{\prime}}$}}
\newcommand{\mdot}{\mbox{$\dot{M}$}}
\newcommand{\msun}{\mbox{${\rm M_{\odot}}$}}
\newcommand{\yr}{\mbox{${\rm yr}^{-1}$}}
\newcommand{\kms}{\mbox{${\rm km} \;{\rm s}^{-1}$}}
\newcommand{\lambdav}{\mbox{${\lambda}_{v}$}}
\newcommand{\lequ}{\mbox{${L_{eq}}$}}
\newcommand{\eqpratio}{\mbox{${R_{eq}/R_{p}}$}}
\newcommand{\ra}{\mbox{${r_{o}}$}}
\newcommand{\bfig}{\begin{figure}[h]}
\newcommand{\efig}{\end{figure}}
\newcommand{\tone}{\mbox{${t_{1}}$}}
\newcommand{\done}{\mbox{${{\rho}_{1}}$}}
\newcommand{\dsn}{\mbox{${\rho}_{SN}$}}
\newcommand{\dzero}{\mbox{${\rho}_{0}$}}
\newcommand{\ve}{\mbox{${v}_{e}$}}
\newcommand{\vej}{\mbox{${v}_{ej}$}}
\newcommand{\Mch}{\mbox{${M}_{ch}$}}
\newcommand{\mej}{\mbox{${M}_{e}$}}
\newcommand{\Mst}{\mbox{${M}_{ST}$}}
\newcommand{\dam}{\mbox{${\rho}_{am}$}}
\newcommand{\Rst}{\mbox{${R}_{ST}$}}
\newcommand{\Vst}{\mbox{${V}_{ST}$}}
\newcommand{\Tst}{\mbox{${T}_{ST}$}}
\newcommand{\no}{\mbox{${n}_{0}$}}
\newcommand{\Efif}{\mbox{${E}_{51}$}}
\newcommand{\rsh}{\mbox{${R}_{sh}$}}
\newcommand{\msh}{\mbox{${M}_{sh}$}}
\newcommand{\vsh}{\mbox{${V}_{sh}$}}
\newcommand{\vrev}{\mbox{${v}_{rev}$}}
\newcommand{\rpr}{\mbox{${R}^{\prime}$}}
\newcommand{\mpr}{\mbox{${M}^{\prime}$}}
\newcommand{\vpr}{\mbox{${V}^{\prime}$}}
\newcommand{\tpr}{\mbox{${t}^{\prime}$}}
\newcommand{\cone}{\mbox{${c}_{1}$}}
\newcommand{\ctwo}{\mbox{${c}_{2}$}}
\newcommand{\cthree}{\mbox{${c}_{3}$}}
\newcommand{\cfour}{\mbox{${c}_{4}$}}
\newcommand{\Te}{\mbox{${T}_{e}$}}
\newcommand{\Ti}{\mbox{${T}_{i}$}}
\newcommand{\Ha}{\mbox{${H}_{\alpha}$}}
\newcommand{\Rprime}{\mbox{${R}^{\prime}$}}
\newcommand{\Vprime}{\mbox{${V}^{\prime}$}}
\newcommand{\Tprime}{\mbox{${T}^{\prime}$}}
\newcommand{\Mprime}{\mbox{${M}^{\prime}$}}
\newcommand{\rprime}{\mbox{${r}^{\prime}$}}
\newcommand{\rfprime}{\mbox{${r}_f^{\prime}$}}
\newcommand{\vprime}{\mbox{${v}^{\prime}$}}
\newcommand{\tprime}{\mbox{${t}^{\prime}$}}
\newcommand{\mprime}{\mbox{${m}^{\prime}$}}
\newcommand{\Me}{\mbox{${M}_{e}$}}
\newcommand{\nh}{\mbox{${n}_{H}$}}
\newcommand{\rr}{\mbox{${R}_{2}$}}
\newcommand{\rf}{\mbox{${R}_{1}$}}
\newcommand{\vtwo}{\mbox{${V}_{2}$}}
\newcommand{\vout}{\mbox{${V}_{1}$}}
\newcommand{\dshell}{\mbox{${{\rho}_{sh}}$}}
\newcommand{\dwind}{\mbox{${{\rho}_{w}}$}}
\newcommand{\dslow}{\mbox{${{\rho}_{s}}$}}
\newcommand{\dfast}{\mbox{${{\rho}_{f}}$}}
\newcommand{\vfast}{\mbox{${v}_{f}$}}
\newcommand{\vslow}{\mbox{${v}_{s}$}}
\newcommand{\cc}{\mbox{${\rm cm}^{-3}$}}
\newcommand{\apj}{\mbox{ApJ}}
\newcommand{\apjl}{\mbox{ApJL}}
\newcommand{\apjs}{\mbox{ApJS}}
\newcommand{\aj}{\mbox{AJ}}
\newcommand{\araa}{\mbox{ARAA}}
\newcommand{\nat}{\mbox{Nature}}
\newcommand{\aap}{\mbox{AA}}
\newcommand{\gca}{\mbox{GeCoA}}
\newcommand{\pasp}{\mbox{PASP}}
\newcommand{\mnras}{\mbox{MNRAS}}
\newcommand{\apss}{\mbox{ApSS}}

\date{}

\pagerange{\pageref{firstpage}--\pageref{lastpage}} \pubyear{2013}

\maketitle

\label{firstpage}

\begin{abstract}

We present newly reduced archival radio observations of \hbox{SN\,1996cr} in
the Circinus Galaxy from the Australia Telescope Compact Array (ATCA) and
the Molonglo Observatory Synthesis Telescope (MOST), and attempt to
model its radio light curves using recent hydrodynamical
simulations of the interaction between the SN ejecta and the
circumstellar material (CSM) at X-ray wavelengths. The radio data within the 
first 1000 days show clear signs of free-free absorption (FFA), which
decreases gradually and is minimal above 1.4~GHz after day $\sim$3000. Constraints
on the FFA optical depth provide estimates of the CSM free electron
density, which allows insight into the ionisation of \hbox{SN\,1996cr}'s
CSM and offers a test on the density distribution adopted by the
hydrodynamical simulation. The intrinsic spectral index of the
radiation shows evidence for spectral flattening, which is characterised
by $\alpha$~=~0.852~$\pm$~0.002 at day 3000 and a decay rate
of $\Delta\alpha$~=~-0.014~$\pm$~0.001 yr$^{-1}$. The striking similarity
in the spectral flattening of SN 1987A, SN 1993J, and SN 1996cr suggests this
may be a relatively common feature of SNe/CSM shocks. We
adopt this spectral index variation to model the
synchrotron radio emission of the shock, and consider several
scalings that relate the parameters of the hydrodynamical simulation to
the magnetic field and electron distribution. The simulated light curves
match the large-scale features of the observed light curves, but fail
to match certain tightly constraining sections. This suggests that
simple energy density scalings may not be able to account for
the complexities of the true physical processes at work, or
alternatively, that the parameters of the simulation require
modification in order to accurately represent the surroundings of \hbox{SN\,1996cr}.
\end{abstract}


\section{Introduction}\label{intro}

Radio and X-ray emission arising from remnants of supernovae (SNe) provide
rich information about the ongoing interaction between the SN ejecta and the
circumstellar material (CSM). The fact that SN ejecta expand $\sim$2--3 orders
of magnitude faster than the slower-moving progenitor stellar wind ensures that,
in a matter of decades, the ejecta will have swept up and revealed
information on tens to thousands of years of stellar evolution leading up
to the explosion, thus providing an alternate window for studying the
evolution and classification of the progenitor stars \citep{Chevalier1994}.

\begin{figure*}
\includegraphics[width=20cm]{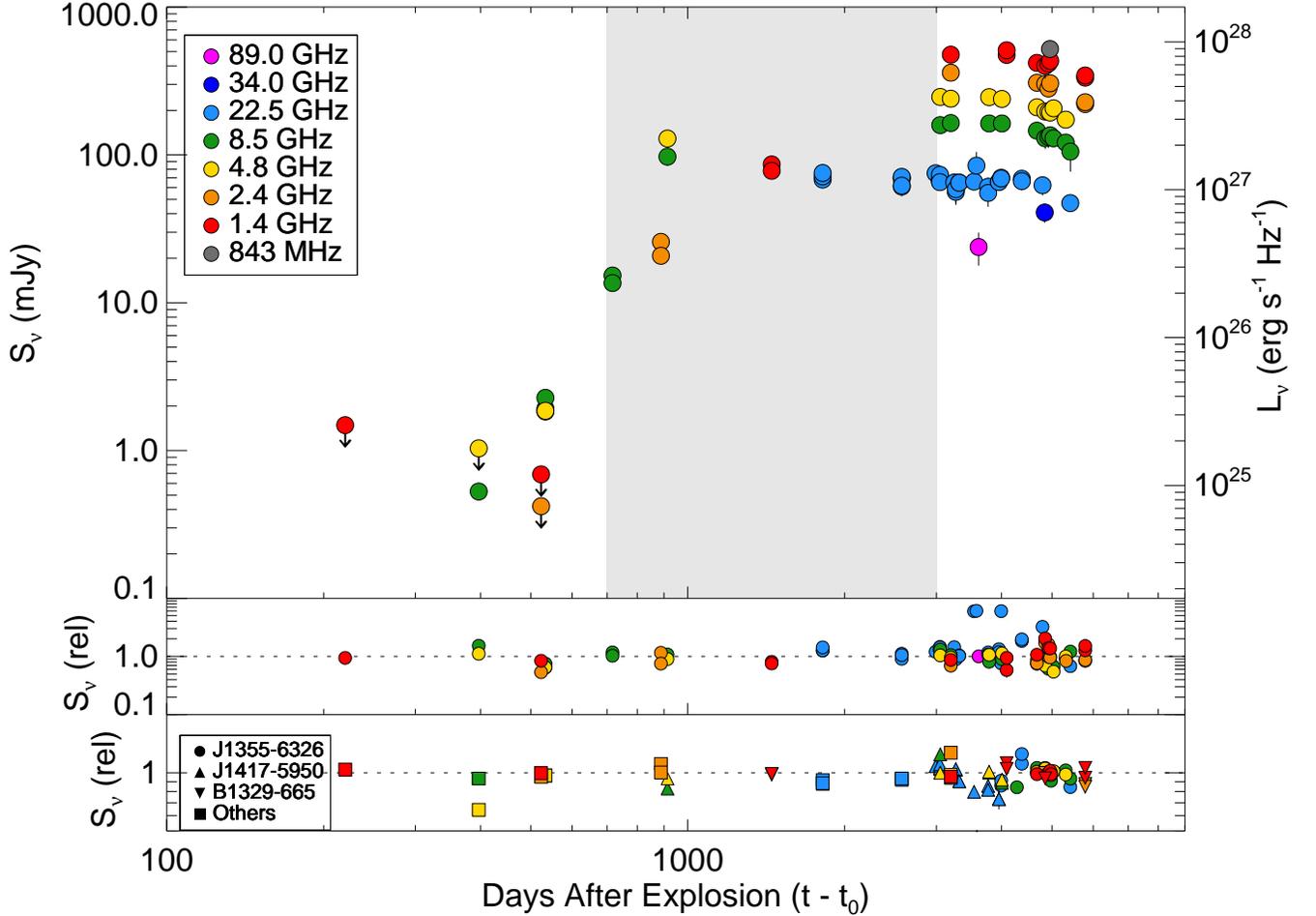}
\caption{Observed radio flux densities (filled circles) for
\hbox{SN\,1996cr} along with 1$\sigma$ error bars (which
are smaller than symbols in some cases); when no secure detection is
obtained we show 3$\sigma$ upper limits (e.g., at early times). The
gray-shaded region denotes the period between days $\sim$700--3000
when the shock is believed to be travelling through the dense
shell. Three regions are clearly defined, with low early emission
(for which only upper limits are available), a flux rise at $\sim$700~d
when the shock begins to interact with the shell, and a late-time plateau
region with a possible turnover at day $\sim$5000. Also shown
are the relative flux densities for the host AGN nucleus (middle panel)
and various phase flux calibrators (lower panel). The nucleus
values were normalised based on the average values at each
frequency found in the highest resolution configurations only
(e.g., 6A), while the calibrators were normalised based on their
historical values at each frequency. In general, the nucleus only
varies by $\approx$10--30$\%$, while the calibrators vary by
$\approx$10--20$\%$. These should provide strict empirical upper
limits to any uncertainties on \hbox{SN\,1996cr}'s variability.}
\label{fig:data_radio}
\end{figure*}

An analytical model of the emission from radio SNe (RSNe) was developed
by \citet[hereafter the `standard' model]{Chevalier1982}, in which the
SN ejecta and CSM density profiles are described as power laws in radius,
and the evolution of the shock front is characterised by a self-similar
solution in which the shock radii expand as power laws in time. The
interaction between the SN ejecta and the CSM leads to the formation
of a contact discontinuity, bounded by a forward shock travelling
into the CSM and a reverse shock moving back into the ejecta
relative to the contact discontinuity. The radio emission arising
from these two shocks is typically characterised as synchrotron
radiation from ultrarelativistic electrons accelerated at the shock
front \citep{Duffy1995, Reynolds2008}. If the energy density of the magnetic 
field as well as the relativistic particles can be assumed as proportional 
to the thermal energy density, then this treatment provides a simple formula
for the radio emission. This has been successful in describing the
radio light curves of SNe exploding into simple red supergiant
(RSG) type winds, where the emission decreases as a power-law in
time \citep{Weiler2002}. It fails, however, to account for the
light curves of RSNe which do not decrease in power-law fashion, 
and therefore where the CSM has presumably been altered by perhaps
a binary companion, or by various evolutionary stages both on
or after the main sequence (cf. table 4 in \citealt{Soderberg2006}). 

The late-stage evolution of stars more massive than $\sim$20M$_\odot$
should extend beyond the initial post-main sequence RSG phase into
yellow or blue supergiant (YSG, BSG), luminous blue variable (LBV),
and/or Wolf-Rayet (WR) phases
(e.g., \citealt{Lamers1991}; \citealt{Stothers1996}). The relatively high
mass-loss rates and/or wind velocities often present during these
later phases (e.g., \citealt{Nugis2000}; \citealt{Vink2011}) can
result in stellar winds with enough momentum to sweep up ambient
material deposited by winds during previous evolutionary stages of the
star, creating a low-density cavity bordered by a dense shell of swept-up material
\citep{Weaver1977}. Once the progenitor explodes, the ejecta expand within
this complex CSM, which is not represented by a power-law density profile, and
for which the standard model, which assumes self-similar shock expansion, is not
applicable (\citealt{Chevalier1989}; \citealt{Tenorio1990, Tenorio1991}; \citealt{Dwarkadas2005}; \citealt{Dwarkadas2007a}). We
{\it must} therefore model the ejecta-CSM interaction through
hydrodynamical simulations using prescribed ejecta and CSM density profiles. The
resulting hydrodynamical quantities (pressure, density and temperature) can
then be input to radiative transfer equations to simulate the radio light curves
and spectra of the SN (e.g., \citealt{Marti2011}; \citealt{Amy2001}).

We employ a simplified version of this method to study the radio
light curve of \hbox{SN\,1996cr}, located in the Circinus
Galaxy ($\sim$3.8 Mpc; \citealt{Freeman1977}; \citealt{Koribalski2004}),
which exploded around 1996 and was confirmed as a type IIn SN
by \citet{Bauer2007} $\sim$11~yr later. Serendipitous archival
data provide a crude picture of the explosion dynamics and the
late evolutionary history of the progenitor (see, e.g., \citealt{Bauer2008}
and references therein). In behaviour reminiscent of
SN 1987A \citep{Park2005, Zanardo2010}, \hbox{SN\,1996cr} shows
an increase of X-ray and radio luminosity in the years following the initial
explosion. \citet{Dwarkadas2010} studied \hbox{SN\,1996cr} through
hydrodynamic simulations of the ejecta-CSM interaction, from
which they accurately reproduced the X-ray light curves and spectra,
and in doing so derived valuable constraints on the CSM
distribution: \hbox{SN\,1996cr} exploded in a low-density medium and
expanded freely before interacting with a dense shell of material at about
0.04 pc, which is thought to be the result of the interaction between fast winds
of either a BSG or WR star and a slower wind, possibly from
a previous RSG stage which is not yet well constrained by observations. 

In this work we provide an observational update on the radio evolution of
\hbox{SN\,1996cr}, and attempt to compute the radio light curves from the
hydrodynamic simulation in \citet{Dwarkadas2010}. The constraints on the
simulation parameters came primarily from the observed X-ray light curves
and spectra. However, there are no X-ray detections prior to year 2000, and 
only a few relatively weak soft X-ray upper limits before that
\citep{Bauer2008, Dwarkadas2010}. Thus for the epoch when the shock is
thought to have first encountered the dense shell in 1997, there are no direct
thermal constraints on the density of the inner cavity implemented in the
simulation. Radio detections of \hbox{SN\,1996cr} start as early as mid-1996, and
could provide constraints on the CSM density during this period, since both
the X-ray and radio emissions can be ultimately related to the density of the
emitting region. Probing the CSM density of the inner boundary of the shell and
possibly even the initial cavity using the radio light curve may ultimately help
to refine some of the uncertainties in the hydrodynamic simulation, give us a better
picture of the CSM structure, and thereby provide more information about the
progenitor star and its evolution.

This study is organised as follows. Radio data and reduction methods are
detailed in \S \ref{radio_obs}. In \S \ref{radio} we characterise the observational
properties of \hbox{SN\,1996cr}, and outline how we translate the hydrodynamic
simulation into a radio emission model. In \S \ref{results} we explore
the H{\sc ii} region surrounding \hbox{SN\,1996cr} and present the simulated
radio light curves. Finally, in \S \ref{discussion} we discuss the results
and implications of this study. In this paper, we adopt the explosion
epoch of 1995.6 determined in \citet{Bauer2008}. All errors are
quoted at 1$\sigma$ unless otherwise stated.

\begin{figure}
\includegraphics[width=8.5cm]{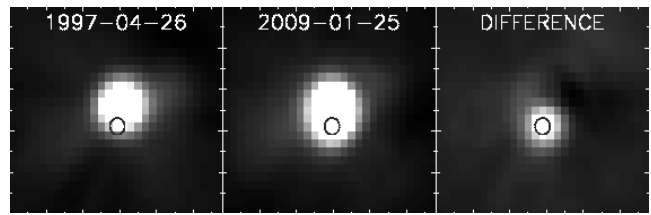}
\caption{We show two 843 MHz images from MOST for the central
4$\farcm$5$\times$4$\farcm$5 of the Circinus Galaxy before (left panel) and
after (middle panel) \hbox{SN\,1996cr} became visible at 843 MHz, as well as a
difference image with the clear \hbox{SN\,1996cr} detection (right panel). A
10$\arcsec$ radius black circle denotes the position of \hbox{SN\,1996cr} in all
panels for reference. All of the images are scaled linearly on an
identical stretch from S$_{\nu}$=0.05--1~Jy. The sidelobes in the images
come from the central gap in the array and do not fully subtract out
due to radio interference.}
\label{fig:sn96cr2_40cm}
\end{figure}

\section{Radio Observations}\label{radio_obs}

\citet{Bauer2008} provide radio data up to 2007. However, since more
data exist for comparison, and some of the data presented in
\citet{Bauer2008} were re-reduced to clean up artifacts and incorporate
a larger array of calibration data, we detail the data reduction
procedures here. The data presented here supersede those of \citet{Bauer2008}.

Numerous observations of the Circinus Galaxy exist from studies at radio
wavelengths spanning 1995 to 2011 (e.g, \citealt{Elmouttie1995}; \citealt{Greenhill1997};
\citealt{Jones1999}; \citealt{McCallum2005,McCallum2009}; \citealt{Phillips1998};
\citealt{Wilson2011}). These were retrieved from the Australia Telescope Compact
Array (ATCA) archive\footnote{http://atoa.atnf.csiro.au/} and were reduced with
{\sc miriad} (v4.0.5 and v20110616) following the procedures outlined in the
ATNF Miriad User Manual\footnote{http://www.atnf.csiro.au/computing/software/miriad/}. Fluxes
for SN\,1996cr are presented in Table~\ref{tab:data_radio} and Fig.~\ref{fig:data_radio}.

For the 1.4, 2.4, 4.8, and 8.5~GHz observations, the `primary' flux
calibrator was typically PKS B1934$-$638, with approximate
flux densities of 14.9, 11.6, 5.8, and 2.8~Jy, respectively. For
15--25~GHz observations the `primary' flux calibrators were
either PKS B1934$-$638 (0.7--1.2~Jy) or PKS B1921$-$293
(10--25 Jy). For the 36~GHz and 89~GHz observations, the primary
calibrators were Mars and Uranus.

The secondary calibrators, which served as the gain, bandpass, and
phase calibrators for the Circinus Galaxy, varied from observation to
observation and were variable in time, and include (in descending
order of use) sources PKS B1921$-$293 (bandpass, alternate 22.5~GHz flux), PKS
B0823$-$500 (bandpass, alternate flux), PMN J1355$-$6326 (phase),
PMN J1417$-$5950 (phase), PKS B1329$-$665 (bandpass, phase),
PKS B1718$-$649 (phase), PKS B1549$-$790 (phase), PMN
J1047$-$6217 (phase), PKS B1104$-$445 (phase), PKS B1424$-$418
(phase), PKS B1610$-$771 (phase), PMN J1342$-$5828 (phase) and
PKS B1236$-$684 (phase). Flux densities for the calibrators are
provided in the Appendix. To cross-check our flux calibration, the
flux densities of the secondary calibrators were compared to their
average or interpolated historical values from calibration monitoring
efforts\footnote{http://www.narrabri.atnf.csiro.au/calibrators/}. Importantly,
aside from one case at day $\sim$400, and in spite of sometimes
strong intrinsic variability and scintillation effects, the phase
calibrators never strayed more than $\approx$20$\%$ from their
nominal historical values, as shown in Fig.~\ref{fig:data_radio}. Both
source and calibration data were inspected with the aim to flag
and remove bad time intervals and channels, time intervals strongly
affected by interference, and obvious emission and absorption
lines. The calibration tables were applied to the Circinus Galaxy
data, from which deconvolved, primary-beam-corrected images
were made. Due to the increasing beam size of ATCA with decreasing frequency,
we measured flux densities using only the 6~km baselines for frequencies
below 4~GHz in order to limit contamination to \hbox{SN\,1996cr} from
the diffuse emission of the host galaxy. Phase
self-calibration was employed when feasible, and only if it led to a
substantial improvement in the rms noise. Point-source flux
densities (determined from {\sc imfit} or {\sc uvfit}) or
\hbox{3$\sigma$} upper limits (determined from {\sc imstat}) were then
determined for \hbox{SN\,1996cr}. These values are provided in
Table~\ref{tab:data_radio} and shown in
Fig.~\ref{fig:data_radio}. The larger errors on the 15--90~GHz data points
reflect greater uncertainties in atmospheric corrections and flux
calibrators. Some of the dispersion among the 15--25~GHz points could
also arise from contamination by weak H$_{2}$O maser emission that was
not flagged from the continuum (e.g., \citealt{Greenhill1997}; \citealt{McCallum2007}).

The only other strong point source in the field of view is the nucleus of
the Circinus Galaxy, which is primarily due to emission from the
Compton-thick AGN as well as a compact circumnuclear starburst. Flux
measurements from the point-like nucleus were extracted in a
manner identical to \hbox{SN\,1996cr}. These are shown in
Fig.~\ref{fig:data_radio} and provided in the Appendix. While
the nucleus could vary intrinsically and can additionally be
contaminated at long wavelengths by diffuse emission from the
extended disk, it at least provides us with a secondary
estimate of any systematic error associated with each observation. We
find that the majority of measurements of the nucleus lie between
10–-30$\%$ of its mean total flux in each band, although some observations
vary upwards considerably. For instance, at 22.5~GHz there may be
possible contamination from water maser emission and varies by a
factor of up to four, while at low frequencies the host galaxy
contaminates as a function of the beamsize and varies by factors as
large as $\sim$10--50. Thus aside from our previously stated reservations
regarding the 15--90~GHz data, we are generally confident
in our calibrated fluxes for \hbox{SN\,1996cr}.

SN 1996cr was observed on 2009~January~25 with the Molonglo Observatory
Synthesis Telescope (MOST) at 843 MHz with a synthesised beam of
47\arcsec\,$\times$\,43\arcsec\, at a position angle of 0$^\circ$. The MOST
data were reduced and calibrated using a custom process described
in \citet{Bock1999}. Given the 24\arcsec north-south separation
of \hbox{SN\,1996cr} from the bright nucleus of the Circinus Galaxy, the
supernova emission was completely blended with that of the galaxy. Fortunately,
there were three prior MOST observations of this field with \hbox{SN\,1996cr}
absent or self-absorbed: 1989 April, 1996 April and 1997 April. Difference
images were formed after regridding all images to a common reference
frame (see Fig.~\ref{fig:sn96cr2_40cm}). A point source fit to each difference
image yielded a mean flux density for \hbox{SN\,1996cr} of 520~$\pm$~30~mJy and
a position fully consistent with the Chandra position. Note that the total flux density for
the Circinus Galaxy in 1996 was $\sim$2.15~$\pm$~0.15~Jy at 843~MHz. Since
the flux contribution from the AGN is expected to be minimal (roughly 75~mJy with
only 10--30$\%$ variability), the observed flux must arise almost exclusively from
diffuse component and hence should not vary.

\section{Modelling the light curve}\label{radio}

\subsection{Radio Emission}\label{radioemission}

The interaction between the SN ejecta and the CSM results in the
formation of a double-shocked structure separated by a contact
discontinuity. Charged particles are accelerated to relativistic
energies across the shock front, giving rise to non-thermal
synchrotron emission in the presence of a strong magnetic
field. Electrons, being substantially less massive, are easier to
accelerate (in the direction perpendicular to their motion) by
the magnetic field, and end up dominating the emission. Their
energy distribution is assumed to be a power law of the form
\hbox{$N(E)$ = $K\,E^{\rm -\gamma}$}, where $E$ is the electron energy, $K$
is the normalisation of the distribution, and $\gamma$ is the electron spectral
index. Following the prescription given in \citet{Chevalier1982}, the 
synchrotron luminosity $L_{\rm \nu}$ at frequency $\nu$ is:

\be
L_{\rm \nu}\,  = \, 4\pi R^2 \Delta R\, K\, B^{\alpha + 1}\,\nu^{-\alpha}\, \rm{e}
^{-\tau}\,\,{\rm (erg\,s^{-1} Hz^{-1})}, \label{eq:lum}
\ee

\noindent where $\Delta R$ is the thickness of the synchrotron-emitting
region at radius $R$, assumed in this case to be the spherically symmetric
region encompassing the forward and reverse shocks, $B$ is the magnetic
field strength, the radio spectral index is $\alpha = (\gamma - 1)/2$,
and $\tau$ is the optical depth. 

\begin{figure}
\hglue-0.8cm{\includegraphics[width=9.5cm]{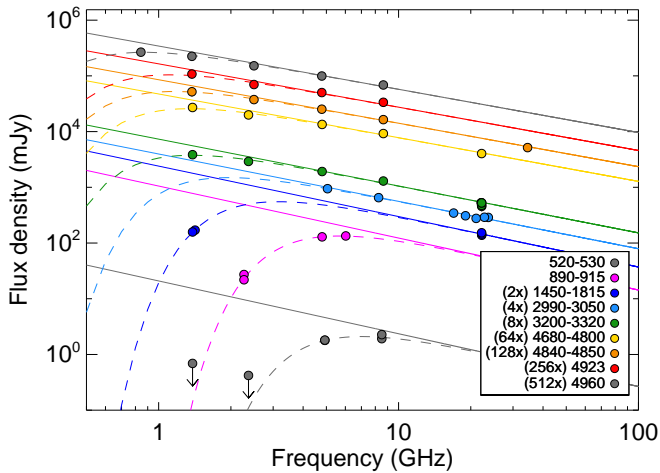}}
\caption{Radio spectra of \hbox{SN\,1996cr} for selected epochs indicated
in the legend as days after explosion. Flux densities (filled circles) are
shown with 1$\sigma$ error bars (which are smaller than symbols in some
cases) or 3$\sigma$ upper limits (downward arrows). The intrinsic
and absorbed spectra are represented by solid and dashed lines
respectively, with the latest spectra at the top. For clarity, fluxes
are plotted with the multiplicative factors in parenthesis. Early-time
spectra show high absorption, mostly affecting the lower frequencies as expected
from FFA. The effect of adopting a varying spectral index (see \S \ref{radio_ppties}
and Fig.~\ref{fig:alpha}) can be seen in the intrinsic models plotted at each epoch.}
\label{fig:spectra}
\end{figure}

Radio emission can be absorbed either by synchrotron self-absorption
(SSA) or free-free absorption (FFA) from CSM both inside (attenuating
backside emission) and outside (attenuating all emission) of the forward
shock, leading to telltale low-frequency absorption. We initially assume all of
the absorption is due to FFA (see Sec.~\ref{radio_ppties} for reasoning) and
use the form of the FFA optical depth given in \citet{Altenhoff1960}:

\be
\tau_{\rm{ff}}\, = \,  8.235\times 10^{-2}\, \nu_{\rm{GHz}}^{-2.1}\, \int_0^{s_{\rm{pc}}} \! {N_e}_{\rm{cm}^{-3}}^2\, T_{\rm{K}}^{-1.35}\, \mathrm{d}s_{\rm{pc}}~, \label{eq:ffa}
\ee

\noindent where $T$ is the temperature of the CSM, $\nu$ is the frequency
and $N_{\rm e}$ is the line-of-sight electron number density. Early absorption
can be high due to the strong density gradient in the stellar wind, which
should decrease as the shock wave overtakes more of the CSM, resulting
in a rapid frequency-dependent turn-on like that observed in the spectra
of \hbox{SN\,1996cr} in Fig.~\ref{fig:spectra}.

\setcounter{table}{0}
\begin{landscape}

\def\arraystretch{1.3}

\begin{deluxetable}{lccc|cc|cc|cc|cc|cc|cc|cc|cc|l}

\tabletypesize{\tiny}
\tablewidth{0pt}
\tablecaption{ATCA Radio Observations\label{tab:data_radio}}
\tablehead{
\colhead{Obs. Date} &
\colhead{Days from} &
\colhead{Obs. ID} &
\colhead{Array} &
\multicolumn{2}{|c|}{W-Band} &
\multicolumn{2}{|c|}{K$_a$-Band} &
\multicolumn{2}{|c|}{K-Band} &
\multicolumn{2}{|c|}{X-Band} &
\multicolumn{2}{|c|}{C-Band} &
\multicolumn{2}{|c|}{S-Band} &
\multicolumn{2}{|c|}{L-Band} &
\multicolumn{2}{|c|}{UHF} \\
& explosion
& 
&
&
\multicolumn{1}{|c}{$\nu$} &
\multicolumn{1}{c|}{$S_{\nu}$} &
\multicolumn{1}{|c}{$\nu$} &
\multicolumn{1}{c|}{$S_{\nu}$} &
\multicolumn{1}{|c}{$\nu$} &
\multicolumn{1}{c|}{$S_{\nu}$} &
\multicolumn{1}{|c}{$\nu$} &
\multicolumn{1}{c|}{$S_{\nu}$} &
\multicolumn{1}{|c}{$\nu$} &
\multicolumn{1}{c|}{$S_{\nu}$} &
\multicolumn{1}{|c}{$\nu$} &
\multicolumn{1}{c|}{$S_{\nu}$} &
\multicolumn{1}{|c}{$\nu$} &
\multicolumn{1}{c|}{$S_{\nu}$} &
 \\
\colhead{(1)}&
\colhead{(2)}&
\colhead{(3)}&
\colhead{(4)}&
\colhead{(5)}&
\colhead{(6)}&
\colhead{(7)}&
\colhead{(8)}&
\colhead{(9)}&
\colhead{(10)}&
\colhead{(11)}&
\colhead{(12)}&
\colhead{(13)}&
\colhead{(14)}&
\colhead{(15)}&
\colhead{(16)}&
\colhead{(17)}&
\colhead{(18)}&
\colhead{(19)}&
\colhead{(20)}&

}

\startdata

1995-03-03 &     -120 & C204a & 750A &     ---     &    ---     &    ---       &    ---     &     ---       &    ---    &         8640 &  $<$  0.3 &         4800 &  $<$  0.6 &     ---     &    ---     &     ---     &    ---     &     ---     &    ---     &     \\
1995-03-03 &     -120 & C204a & 750A &     ---     &    ---     &    ---       &    ---     &     ---       &    ---    &         8640 &  $<$  0.3 &         4800 &  $<$  0.4 &     ---     &    ---     &     ---     &    ---     &     ---     &    ---     &     \\
1995-03-30 &      -93 & C363 & 1.5A &     ---     &    ---     &    ---       &    ---     &     ---       &    ---    &     ---      &    ---     &     ---     &    ---     &     ---     &    ---     &         1418 &  $<$  6.6 &     ---     &    ---     &     \\
1995-04-16 &      -76 & C418 & 6C &     ---     &    ---     &    ---       &    ---     &     ---       &    ---    &     ---      &    ---     &     ---     &    ---     &         2768 &  $<$  1.1 &         1418 &  $<$  1.2 &     ---     &    ---     &     \\
1995-05-27 &      -35 & C204a & 1.5B &     ---     &    ---     &    ---       &    ---     &     ---       &    ---    &         8640 &  $<$  0.3 &         4800 &  $<$  0.2 &     ---     &    ---     &     ---     &    ---     &     ---     &    ---     &     \\
1995-05-27 &      -35 & C204a & 1.5B &     ---     &    ---     &    ---       &    ---     &     ---       &    ---    &         8640 &  $<$  0.3 &         4800 &  $<$  0.5 &     ---     &    ---     &     ---     &    ---     &     ---     &    ---     &     \\
1995-06-07 &      -24 & C204a & 750C &     ---     &    ---     &    ---       &    ---     &     ---       &    ---    &         8640 &  $<$  0.8 &         4800 &  $<$  0.3 &     ---     &    ---     &     ---     &    ---     &     ---     &    ---     &     \\
1995-06-07 &      -24 & C204a & 750C &     ---     &    ---     &    ---       &    ---     &     ---       &    ---    &         8640 &  $<$  0.3 &         4800 &  $<$  0.2 &     ---     &    ---     &     ---     &    ---     &     ---     &    ---     &     \\
1995-07-27 &       26 & C204c & 6C &     ---     &    ---     &    ---       &    ---     &     ---       &    ---    &     ---      &    ---     &     ---     &    ---     &         2368 &  $<$  1.3 &         1376 &  $<$  3.4 &     ---     &    ---     &     \\
1995-08-03 &       33 & C363 & 375 &     ---     &    ---     &    ---       &    ---     &     ---       &    ---    &     ---      &    ---     &     ---     &    ---     &     ---     &    ---     &         1418 &  $<$ 16.7 &     ---     &    ---     &     \\
1996-02-06 &      220 & CT09 & 750B &     ---     &    ---     &    ---       &    ---     &     ---       &    ---    &     ---      &    ---     &     ---     &    ---     &     ---     &    ---     &         1664 &  $<$  1.5 &     ---     &    ---     &     \\
1996-08-01 $^{(1)}$&      397 & C466 & 6C &     ---     &    ---     &    ---       &    ---     &     ---       &    ---    &         8640 &   0.5 $\pm$  0.0 &         4800 &  $<$  1.0 &     ---     &    ---     &     ---     &    ---     &     ---     &    ---     &     \\
1996-12-05 $^{(2)}$&      523 & C505 & 365 &     ---     &    ---     &    ---       &    ---     &     ---       &    ---    &     ---      &    ---     &     ---     &    ---     &         2368 &  $<$  0.4 &         1384 &  $<$  0.7 &     ---     &    ---     &     \\
1996-12-15 $^{(2)}$&      533 & C586 & 6D &     ---     &    ---     &    ---       &    ---     &     ---       &    ---    &         8512 &   1.9 $\pm$  0.1 &         4928 &   1.8 $\pm$  0.1 &     ---     &    ---     &     ---     &    ---     &     ---     &    ---     &     \\
1996-12-15 $^{(2)}$&      533 & C586 & 6D &     ---     &    ---     &    ---       &    ---     &     ---       &    ---    &         8512 &   2.3 $\pm$  0.1 &         4928 &   1.8 $\pm$  0.1 &     ---     &    ---     &     ---     &    ---     &     ---     &    ---     &     \\
1997-06-17 &      717 & V100C & 6A &     ---     &    ---     &    ---       &    ---     &     ---       &    ---    &         8425 &  15.4 $\pm$  0.3 &     ---     &    ---     &     ---     &    ---     &     ---     &    ---     &     ---     &    ---     &     \\
1997-06-17 &      717 & V100C & 6A &     ---     &    ---     &    ---       &    ---     &     ---       &    ---    &         8425 &  13.7 $\pm$  0.3 &     ---     &    ---     &     ---     &    ---     &     ---     &    ---     &     ---     &    ---     &     \\
1997-12-05 $^{(3)}$&      888 & V099 & 6C &     ---     &    ---     &    ---       &    ---     &     ---       &    ---    &     ---      &    ---     &     ---     &    ---     &         2268 &  27.2 $\pm$  2.1 &     ---     &    ---     &     ---     &    ---     &     \\
1997-12-05 $^{(3)}$&      888 & V099 & 6C &     ---     &    ---     &    ---       &    ---     &     ---       &    ---    &     ---      &    ---     &     ---     &    ---     &         2268 &  21.9 $\pm$  1.1 &     ---     &    ---     &     ---     &    ---     &     \\
1997-12-31 $^{(3)}$&      914 & C694 & 6C &     ---     &    ---     &    ---       &    ---     &     ---       &    ---    &         6024 & 134.0 $\pm$ 13.1 &         4800 & 129.0 $\pm$  8.6 &     ---     &    ---     &     ---     &    ---     &     ---     &    ---     &     \\
1999-06-19 $^{(4)}$&     1449 & C788 & 375 &     ---     &    ---     &    ---       &    ---     &     ---       &    ---    &     ---      &    ---     &     ---     &    ---     &     ---     &    ---     &         1418 &  85.2 $\pm$  2.1 &     ---     &    ---     &     \\
1999-06-19 $^{(4)}$&     1449 & C788 & 375 &     ---     &    ---     &    ---       &    ---     &     ---       &    ---    &     ---      &    ---     &     ---     &    ---     &     ---     &    ---     &         1384 &  78.9 $\pm$  1.5 &     ---     &    ---     &     \\
2000-06-19 $^{(4)}$&     1815 & V137 & 6B &     ---     &    ---     &    ---       &    ---     &        22215 &  68.9 $\pm$  8.5 &     ---      &    ---     &     ---     &    ---     &     ---     &    ---     &     ---     &    ---     &     ---     &    ---     &     \\
2000-06-19 $^{(4)}$&     1815 & V137 & 6B &     ---     &    ---     &    ---       &    ---     &        22203 &  72.4 $\pm$  9.1 &     ---      &    ---     &     ---     &    ---     &     ---     &    ---     &     ---     &    ---     &     ---     &    ---     &     \\
2000-06-19 $^{(4)}$&     1815 & V137 & 6B &     ---     &    ---     &    ---       &    ---     &        22190 &  76.4 $\pm$ 10.0 &     ---      &    ---     &     ---     &    ---     &     ---     &    ---     &     ---     &    ---     &     ---     &    ---     &     \\
2002-07-18 &     2574 & V137 & 1.5G &     ---     &    ---     &    ---       &    ---     &        22224 &  69.5 $\pm$  8.8 &     ---      &    ---     &     ---     &    ---     &     ---     &    ---     &     ---     &    ---     &     ---     &    ---     &     \\
2002-07-18 &     2574 & V137 & 1.5G &     ---     &    ---     &    ---       &    ---     &        22216 &  70.8 $\pm$  5.6 &     ---      &    ---     &     ---     &    ---     &     ---     &    ---     &     ---     &    ---     &     ---     &    ---     &     \\
2002-07-18 &     2574 & V137 & 1.5G &     ---     &    ---     &    ---       &    ---     &        22201 &  61.8 $\pm$  8.7 &     ---      &    ---     &     ---     &    ---     &     ---     &    ---     &     ---     &    ---     &     ---     &    ---     &     \\
2002-07-18 &     2574 & V137 & 1.5G &     ---     &    ---     &    ---       &    ---     &        22187 &  71.9 $\pm$  8.6 &     ---      &    ---     &     ---     &    ---     &     ---     &    ---     &     ---     &    ---     &     ---     &    ---     &     \\
2002-07-18 &     2574 & V137 & 1.5G &     ---     &    ---     &    ---       &    ---     &        22172 &  62.7 $\pm$  8.2 &     ---      &    ---     &     ---     &    ---     &     ---     &    ---     &     ---     &    ---     &     ---     &    ---     &     \\
2003-09-07 $^{(5)}$&     2990 & C1224 & EW367 &     ---     &    ---     &    ---       &    ---     &        23659 &  72.0 $\pm$  7.4 &     ---      &    ---     &     ---     &    ---     &     ---     &    ---     &     ---     &    ---     &     ---     &    ---     &     \\
2003-11-03 $^{(5)}$&     3047 & C1049 & H214 &     ---     &    ---     &    ---       &    ---     &        19008 &  77.1 $\pm$  7.8 &     ---      &    ---     &     ---     &    ---     &     ---     &    ---     &     ---     &    ---     &     ---     &    ---     &     \\
2003-11-03 $^{(5)}$&     3047 & C1049 & H214 &     ---     &    ---     &    ---       &    ---     &        16960 &  86.2 $\pm$  9.4 &     ---      &    ---     &     ---     &    ---     &     ---     &    ---     &     ---     &    ---     &     ---     &    ---     &     \\
2003-11-04 $^{(5)}$&     3048 & C1049 & 1.5D &     ---     &    ---     &    ---       &    ---     &        22796 &  72.5 $\pm$  7.1 &     ---      &    ---     &     ---     &    ---     &     ---     &    ---     &     ---     &    ---     &     ---     &    ---     &     \\
2003-11-04 $^{(5)}$&     3048 & C1049 & 1.5D &     ---     &    ---     &    ---       &    ---     &        21056 &  69.2 $\pm$  7.4 &     ---      &    ---     &     ---     &    ---     &     ---     &    ---     &     ---     &    ---     &     ---     &    ---     &     \\
2003-11-08 $^{(5)}$&     3052 & C1049 & 1.5D &     ---     &    ---     &    ---       &    ---     &     ---       &    ---    &         8256 & 163.0 $\pm$  7.3 &         5056 & 236.0 $\pm$  4.5 &     ---     &    ---     &     ---     &    ---     &     ---     &    ---     &     \\
2004-04-02 $^{(6)}$&     3198 & C1424 & 6A &     ---     &    ---     &    ---       &    ---     &     ---       &    ---    &         8640 & 162.0 $\pm$  1.3 &         4800 & 240.0 $\pm$  1.1 &         2368 & 364.0 $\pm$  2.7 &         1384 & 481.0 $\pm$  7.5 &     ---     &    ---     &     \\
2004-05-20 $^{(6)}$&     3246 & CX065 & 1.5B &     ---     &    ---     &    ---       &    ---     &        22087 &  66.1 $\pm$  9.2 &     ---      &    ---     &     ---     &    ---     &     ---     &    ---     &     ---     &    ---     &     ---     &    ---     &     \\
2004-06-13 $^{(6)}$&     3270 & V176 & 750D &     ---     &    ---     &    ---       &    ---     &        22209 &  56.8 $\pm$ 10.3 &     ---      &    ---     &     ---     &    ---     &     ---     &    ---     &     ---     &    ---     &     ---     &    ---     &     \\
2004-06-13 $^{(6)}$&     3270 & V176 & 750D &     ---     &    ---     &    ---       &    ---     &        22195 &  59.3 $\pm$ 11.7 &     ---      &    ---     &     ---     &    ---     &     ---     &    ---     &     ---     &    ---     &     ---     &    ---     &     \\
2004-08-01 $^{(6)}$&     3319 & V176B & H168 &     ---     &    ---     &    ---       &    ---     &        22209 &  65.7 $\pm$  8.8 &     ---      &    ---     &     ---     &    ---     &     ---     &    ---     &     ---     &    ---     &     ---     &    ---     &     \\
2004-08-01 $^{(6)}$&     3319 & V176B & H168 &     ---     &    ---     &    ---       &    ---     &        22195 &  65.6 $\pm$  8.1 &     ---      &    ---     &     ---     &    ---     &     ---     &    ---     &     ---     &    ---     &     ---     &    ---     &     \\
2005-03-14 $^{(7)}$&     3544 & C1368 & H214 &     ---     &    ---     &    ---       &    ---     &        22193 &  66.7 $\pm$  7.8 &     ---      &    ---     &     ---     &    ---     &     ---     &    ---     &     ---     &    ---     &     ---     &    ---     &     \\
2005-04-20 $^{(7)}$&     3581 & C1368 & 750A &     ---     &    ---     &    ---       &    ---     &        22194 &  85.6 $\pm$ 20.1 &     ---      &    ---     &     ---     &    ---     &     ---     &    ---     &     ---     &    ---     &     ---     &    ---     &     \\
2005-05-25 $^{(7)}$&     3616 & C1321 & Hybrid &        88632 &  23.9 $\pm$  6.0 &    ---       &    ---     &     ---       &    ---    &     ---      &    ---     &     ---     &    ---     &     ---     &    ---     &     ---     &    ---     &     ---     &    ---     &     \\
2005-10-27 $^{(7)}$&     3771 & C1049 & H168 &     ---     &    ---     &    ---       &    ---     &        18752 &  70.8 $\pm$  9.2 &     ---      &    ---     &     ---     &    ---     &     ---     &    ---     &     ---     &    ---     &     ---     &    ---     &     \\
2005-10-27 $^{(7)}$&     3771 & C1049 & H168 &     ---     &    ---     &    ---       &    ---     &        21056 &  58.4 $\pm$ 10.8 &     ---      &    ---     &     ---     &    ---     &     ---     &    ---     &     ---     &    ---     &     ---     &    ---     &     \\
2005-11-13 $^{(7)}$&     3788 & C1049 & 1.5C &     ---     &    ---     &    ---       &    ---     &     ---       &    ---    &         8640 & 161.0 $\pm$  2.2 &         4800 & 246.0 $\pm$  1.7 &     ---     &    ---     &     ---     &    ---     &     ---     &    ---     &     \\
2006-04-29 $^{(8)}$&     3955 & C1049 & H214 &     ---     &    ---     &    ---       &    ---     &        18752 &  76.4 $\pm$  7.3 &     ---      &    ---     &     ---     &    ---     &     ---     &    ---     &     ---     &    ---     &     ---     &    ---     &     \\
2006-04-29 $^{(8)}$&     3955 & C1049 & H214 &     ---     &    ---     &    ---       &    ---     &        21056 &  69.2 $\pm$  7.3 &     ---      &    ---     &     ---     &    ---     &     ---     &    ---     &     ---     &    ---     &     ---     &    ---     &     \\
2006-06-08 $^{(8)}$&     3995 & CX114 & 1.5D &     ---     &    ---     &    ---       &    ---     &        22193 &  70.8 $\pm$  9.0 &     ---      &    ---     &     ---     &    ---     &     ---     &    ---     &     ---     &    ---     &     ---     &    ---     &     \\
2006-06-08 $^{(8)}$&     3995 & CX114 & 1.5D &     ---     &    ---     &    ---       &    ---     &        20160 &  75.7 $\pm$  7.6 &     ---      &    ---     &     ---     &    ---     &     ---     &    ---     &     ---     &    ---     &     ---     &    ---     &     \\
2006-06-21 $^{(8)}$&     4008 & C1049 & 1.5D &     ---     &    ---     &    ---       &    ---     &     ---       &    ---    &         8640 & 161.0 $\pm$  7.1 &         4800 & 239.0 $\pm$  6.0 &     ---     &    ---     &     ---     &    ---     &     ---     &    ---     &     \\
2006-09-14 $^{(8)}$&     4093 & C1341 & H75 &     ---     &    ---     &    ---       &    ---     &     ---       &    ---    &     ---      &    ---     &     ---     &    ---     &     ---     &    ---     &         1384 & 479.0 $\pm$  2.5 &     ---     &    ---     &     \\
2006-09-14 $^{(8)}$&     4093 & C1341 & H75 &     ---     &    ---     &    ---       &    ---     &     ---       &    ---    &     ---      &    ---     &     ---     &    ---     &     ---     &    ---     &         1418 & 505.0 $\pm$  3.2 &     ---     &    ---     &     \\
2007-06-24 &     4376 & VX013A & EW352 &     ---     &    ---     &    ---       &    ---     &        22316 &  69.4 $\pm$  7.7 &     ---      &    ---     &     ---     &    ---     &     ---     &    ---     &     ---     &    ---     &     ---     &    ---     &     \\
2007-06-24 &     4376 & VX013A & EW352 &     ---     &    ---     &    ---       &    ---     &        22348 &  66.8 $\pm$  9.0 &     ---      &    ---     &     ---     &    ---     &     ---     &    ---     &     ---     &    ---     &     ---     &    ---     &     \\
2008-04-21 $^{(9)}$&     4678 & C1803 & 6A &     ---     &    ---     &    ---       &    ---     &     ---       &    ---    &         8640 & 144.0 $\pm$  4.0 &         4800 & 210.0 $\pm$  4.0 &         2368 & 311.0 $\pm$ 11.0 &         1384 & 423.0 $\pm$ 27.3 &     ---     &    ---     &     \\
2008-08-16 $^{(9)}$&     4795 & C1750 & 6B &     ---     &    ---     &    ---       &    ---     &        22192 &  63.0 $\pm$  9.0 &     ---      &    ---     &     ---     &    ---     &     ---     &    ---     &     ---     &    ---     &     ---     &    ---     &     \\
2008-10-02 $^{(10)}$&     4842 & C184 & H168 &     ---     &    ---     &        34496 &  40.3 $\pm$  6.0 &     ---       &    ---    &     ---      &    ---     &     ---     &    ---     &     ---     &    ---     &     ---     &    ---     &     ---     &    ---     &     \\
2008-10-12 $^{(10)}$&     4852 & C184 & 6A &     ---     &    ---     &    ---       &    ---     &     ---       &    ---    &         8640 & 128.0 $\pm$ 20.0 &         4786 & 197.0 $\pm$ 20.1 &     ---     &    ---     &     ---     &    ---     &     ---     &    ---     &     \\
2008-10-12 $^{(10)}$&     4852 & C184 & 6A &     ---     &    ---     &    ---       &    ---     &     ---       &    ---    &     ---      &    ---     &     ---     &    ---     &         2496 & 293.0 $\pm$ 39.8 &         1376 & 405.0 $\pm$ 71.0 &     ---     &    ---     &     \\
2008-12-22 $^{(11)}$&     4923 & C184 & 6C &     ---     &    ---     &    ---       &    ---     &     ---       &    ---    &         8640 & 131.0 $\pm$ 21.0 &         4786 & 196.0 $\pm$ 19.0 &         2496 & 273.0 $\pm$ 23.0 &         1376 & 421.0 $\pm$ 65.0 &     ---     &    ---     &     \\
2009-01-25 $^{(12)}$&     4957 & - & MOST &     ---     &    ---     &    ---       &    ---     &     ---       &    ---    &     ---      &    ---     &     ---     &    ---     &     ---     &    ---     &     ---     &    ---     &          843 & 520.0 $\pm$ 30.0 &     \\
2009-01-28 $^{(12)}$&     4960 & C184 & 1.5C &     ---     &    ---     &    ---       &    ---     &     ---       &    ---    &         8640 & 134.0 $\pm$ 10.3 &         4786 & 194.0 $\pm$ 13.0 &         2496 & 296.0 $\pm$ 32.0 &         1376 & 438.0 $\pm$ 60.0 &     ---     &    ---     &     \\
2009-04-13 $^{(13)}$&     5035 & CX172 & H168 &     ---     &    ---     &    ---       &    ---     &     ---       &    ---    &         8800 & 126.0 $\pm$  9.1 &         5500 & 186.0 $\pm$  8.9 &     ---     &    ---     &     ---     &    ---     &     ---     &    ---     &     \\
2010-01-16 $^{(14)}$&     5313 & C1726 & 6A &     ---     &    ---     &    ---       &    ---     &     ---       &    ---    &         9000 & 116.0 $\pm$ 10.7 &         5500 & 156.0 $\pm$  9.9 &     ---     &    ---     &     ---     &    ---     &     ---     &    ---     &     \\
2010-05-05 $^{(14)}$&     5422 & V253 & H214 &     ---     &    ---     &    ---       &    ---     &        22000 &  48.0 $\pm$  5.3 &     ---      &    ---     &     ---     &    ---     &     ---     &    ---     &     ---     &    ---     &     ---     &    ---     &     \\
2010-05-07 $^{(14)}$&     5424 & V253 & H214 &     ---     &    ---     &    ---       &    ---     &     ---       &    ---    &         8425 & 106.0 $\pm$ 28.1 &     ---     &    ---     &     ---     &    ---     &     ---     &    ---     &     ---     &    ---     &     \\
2011-05-09 &     5791 & C2479 & 1.5B &     ---     &    ---     &    ---       &    ---     &     ---       &    ---    &     ---      &    ---     &     ---     &    ---     &         2100 & 244.0 $\pm$  8.5 &         1418 & 330.0 $\pm$ 40.1 &     ---     &    ---     &     \\
2011-05-09 &     5791 & C2479 & 1.5B &     ---     &    ---     &    ---       &    ---     &     ---       &    ---    &     ---      &    ---     &     ---     &    ---     &         2100 & 251.0 $\pm$  6.3 &         1613 & 310.0 $\pm$ 29.9 &     ---     &    ---     &     \\

\enddata
\begin{flushleft}

\tablecomments{
{\it Column 1:} Starting date of observation. Numbers in
parenthesis indicate epochs whose data were combined,
as described  in Sec.~\ref{radio_ppties}.
{\it Column 2:} Days from explosion date.
{\it Column 3:} Observing program ID.
{\it Column 4:} Array configuration.
{\it Columns 5, 7, 9, 11, 13, 15, 17, and 19:} Mean frequency of observed band in
units of MHz.
{\it Columns 6, 8, 10, 12, 14, 16, 18, and 20:} Integrated flux densities, as
determined from {\sc imfit}, or \hbox{3$\sigma$} upper limits, as
determined from {\sc imstat}, in units of mJy. Uncertainties include both
statistical and systematic error terms. The systematic error is estimated
from the ratio of the measured calibrator fluxes over its estimated
historical value based on monitored light curves (typical variance was
5--20\%).
}
\end{flushleft}

\end{deluxetable}
\end{landscape}

\subsection{Empirical Radio Properties}\label{radio_ppties}

A key constraint on the nature of the radio emission is its spectral
index. Measuring this for \hbox{SN\,1996cr}, however, is non-trivial, since
this object was largely observed via serendipitous observations with
sparse frequency coverage, while additionally the data during the first
1000 days appear to be heavily absorbed (Fig.~\ref{fig:spectra}). In
order to deal with the lack of multifrequency observations at each
epoch, the available data were grouped together in periods of up to
six months where possible, during which the spectral index is assumed
to remain constant to within our measurement errors. 

A constant spectral index is generally assumed for radio SNe, but the data
for \hbox{SN\,1996cr} allow us to confront this assumption directly. We
note that the late-time data for \hbox{SN\,1996cr}, between days
$\sim$2800--5000, exhibit minimal absorption, and thus provide relatively
strong constraints on the intrinsic spectral index. For epochs with wide
frequency coverage, an FFA optical depth and spectral index were computed
using the Levenberg-Marquart least-squares method. Limits on the optical
depth for epochs with poor low-frequency coverage were set equal to the
value from the last well-fit spectrum. These limits are of minimal
consequence, as the free-fit spectral indices in epochs where such
optical depth limits exist only show marginal variations
to changes in the limiting optical depth values. The constraints on the optical depths
and the values of the spectral indices are listed in Table~\ref{tab:constraints_tau}.

\begin{figure}
\hglue-0.8cm{\includegraphics[width=9.5cm]{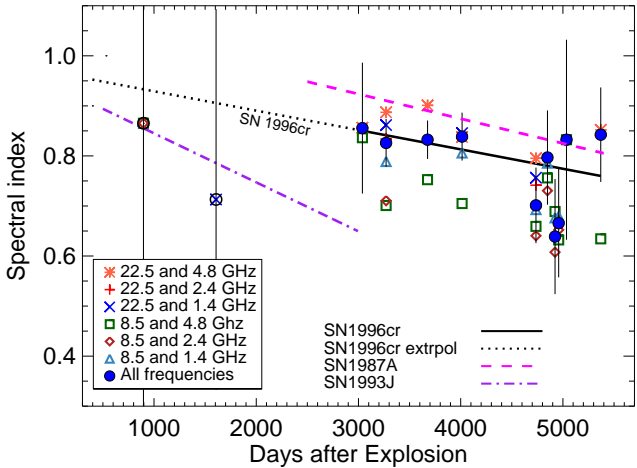}}
\caption{Points show the intrinsic spectral indices and 1$\sigma$
errors derived from linear regression fits to the \hbox{SN\,1996cr}
spectra shown in Fig.~\ref{fig:spectra}. Also shown are all
available two-frequency spectral indices for a given epoch (errors are
not shown for plot clarity, but can be assessed using
Table~\ref{tab:data_radio}). A linear
regression fit to the well constrained spectral indices from
days 3000--5000 versus time suggests that the spectral slope may be
flattening, similar to the spectral index variations seen in both SN 1987A
and SN 1993J (see text for details), although the result is not
yet robust. Early best-fitting spectral indices for \hbox{SN\,1996cr} are not
included in this parametrisation, but are consistent with our extrapolation 
of the best-fitting time evolution to within their large errors.}
\label{fig:alpha}
\end{figure}

As seen in Fig.~\ref{fig:alpha}, well constrained values of the intrinsic
spectral index between days $\sim$3000--5000 appear to shift to
flatter values at later times, and a weighted linear regression fit (OLS Y$|$X; \citealt{Isobe1990})
to the values of the spectral index shows evidence for
mild spectral flattening (nominally $\sim$10$\sigma$ confidence), which we
assume to be the intrinsic variation of the spectral index. A constant
alpha model, however, cannot be ruled out using $\chi^{2}$ or likelihood
ratio statistics (only $\sim$1.5$\sigma$ confidence), leaving doubts about the
veracity of the trend. Extrapolation of this trend back to days $\sim$520,
$\sim$900 and $\sim$1600--1900 is consistent with the data, although
these values are not well constrained due to stronger degeneracies
between their spectral indices and absorption.

We characterise the evolution of the intrinsic spectral index of
\hbox{SN\,1996cr} in a manner similar to that used for
\hbox{SN\,1987A} in \citet{Zanardo2010}. The intrinsic spectral index
is given by \hbox{$\alpha(t)$~=~$\alpha_0$~$+$~$\beta_0$~$\times$~($t-t_0$)/$\Delta$},
where $t$ is given in days, $t_0$~=~3000~d,  $\Delta$~=~365.25~d,  
$\alpha_0$~=~0.852~$\pm$~0.002, and \hbox{$\beta_0$~=~-0.014~$\pm$~0.001}. Errors
on regression parameters were assessed through jackknife resampling. We
extrapolate this regression back to day 400 (see Fig.~\ref{fig:alpha})
and rely on the fixed values of the spectral index from the above
fit to reassess values of the optical depth that will be used in the
subsequent analysis. The specific (fixed) values of the spectral index
and the refitted values of the optical depth are also listed in
Table~\ref{tab:constraints_tau}.

The mild evolution of the spectral index of \hbox{SN\,1996cr} is
similar to that of the highly monitored \hbox{SN\,1987A}
\citep{Zanardo2010}. Fig.~\ref{fig:alpha} also shows the
spectral index variation for \hbox{SN\,1993J}; we note that
\citet{Weiler2007} assumed a constant spectral index for this SN, but
its spectral index clearly appears to vary based on fig. 4 of
\citet{Weiler2007}, fig. 7 of \citet{Chandra2004}, and fig. 8 of
\citet{Marti2011}. From the data in the $\sim$80--3000~d period from
\citet{Weiler2007}, a constant spectral model is rejected
at a 100$\%$ confidence level. We characterise a varying spectral
model for \hbox{SN\,1993J} as \hbox{$\alpha$~=~0.865~$\pm$~0.003}
at day 800, and a decay rate of \hbox{$\Delta\alpha$~=~-0.036~$\pm$~0.001 yr$^{-1}$}. 

It is intriguing that the spectral indices of arguably the three best-studied
radio SNe all show evidence for spectral flattening with time. This variation
runs counter to some theoretical predictions (e.g., \citealt{Ellison2004};
\citealt{Bere2006}; \citealt{Tati2009}) and clearly warrants further
investigation. It is still unclear whether this spectral flattening is a ubiquitous
property of young RSNe. For instance, SN 1979C also has a considerable amount
of radio data available and appears to remain constant since day 2000 (see
fig. 3 of \citealt{Montes2000}), although difficulties associated with disentangling
spectral and CSM variations and the large scatter in measured spectral
indices could easily mask subtle spectral changes such as those observed in
Fig.~\ref{fig:alpha}. SN 1986J, on the other hand, shows strong, complex
spectral variations with time. VLBI observations suggest that these variations
appear to stem from the flat spectrum of the remnant core brightening as
the steep spectrum of the shell fades \citep{Bieten2010}. Thus neither
SN 1979C nor SN 1986J provide `clean' examples. Other RSNe, such as SN 1980K
and SN 1994I show hints of spectral variations, while SN 1978K, SN 1995N
and SN2001ig remain thoroughly ambiguous
\citep{Montes1998, Weiler2011, Smith2007, Chandra2009, Ryder2004}; unfortunately,
all are relatively weak radio sources and hence too poorly
constrained to provide compelling examples either for or against spectral
flattening. Clearly interpreting mild evolutionary trends (e.g., SN 1993J)
should be done with caution, given the obvious degeneracies between
parameters responsible for the synchrotron spectrum in SNe. Past and
future LBA imaging can marginally resolve \hbox{SN\,1996cr} and could
highlight any strong spectral variations in this source (Bartel et al., in preparation).

The brightness temperature of \hbox{SN\,1996cr}, which is calculated from
the simulated shock expansion and the FFA-corrected data, does not exceed
the value of $\approx$2--4~$\times$~10$^{11}$~K (there is a very weak
frequency dependence) which typically leads to SSA
\citep{Readhead1994}. However, this limit may be reached for hypothetical low
frequency data points around day 800, which suggests a possible combination
of FFA and SSA at early times, although this is not well constrained with our
limited data. We model this in more detail in the next section.

\begin{table*}
\caption{Selected epochs with corresponding free-fit optical depth and
spectral index, final optical depth, and fixed spectral
index.\label{tab:constraints_tau}} 
\centering 
\begin{tabular}{c c c c c} 
\hline\hline Days since explosion & $\tau_0$ (K$^{-1.35}$\,pc\,cm$^{-6}$) & Spectral Index & $\tau_0$ (K$^{-1.35}$\,pc\,cm$^{-6}$) & Spectral Index\\ [0.5ex]	
 & (free-fit) & (free-fit) & (refitted) & (fixed)\\ [0.5ex]	
\hline

397 $^{ }$ &     	30.00~$\dagger$		&     1.00~$\dagger$ 		&     	30.00~$\dagger$ 	&    0.95\\ 
530$^{*}$ &     	28.16~$\pm$~2.56 	&     1.00~$\dagger$ 		&     	27.07~$\pm$~2.54 &    0.95\\ 
897$^{*}$ &     	16.67~$\pm$~4.97 	&	   0.87~$\pm$~0.86 	&    	17.05~$\pm$~0.51 &    0.93\\ 
1607$^{*}$ &     	3.75~$\pm$~2.11 		&    0.71~$\pm$~0.38		&   	4.82~$\pm$~0.16 	&    0.91\\ 
3039$^{*}$ &     	1.87~$\pm$~5.20 		&    0.86~$\pm$~0.13 	&   	1.66~$\pm$~1.60 	&    0.85\\ 
3270$^{*}$ &    	0.72~$\pm$~0.23 	&    0.83~$\pm$~0.05 	&    	0.78~$\pm$~0.13 	&    0.84\\ 
3676 $^{ }$&    	0.70~$\dagger$ 		&    0.83~$\pm$~0.04 	&    	0.70~$\dagger$ 		&    0.83\\ 
4013 $^{ }$&    	0.73~$\pm$~0.18 	&    0.84~$\pm$~0.05 	&    	0.64~$\pm$~0.06 	&    0.81\\ 
4737$^{*}$ &    	0.39~$\pm$~0.34 	&    0.70~$\pm$~0.08 	&    	0.70~$\pm$~0.19 	&    0.79\\ 
4849$^{*}$ &    	0.58~$\pm$~0.58 	&    0.80~$\pm$~0.09 	&   	0.51~$\pm$~0.40 	&    0.78\\ 
4923$^{*}$ &    	0.15~$\dagger$		&    0.64~$\pm$~0.12 	&    	0.46~$\pm$~0.41 	&    0.78\\ 
4959$^{*}$ &    	0.13~$\pm$~0.15 	&    0.67~$\pm$~0.11 	&    	0.27~$\pm$~0.06 	&    0.78\\ 
5035 $^{ }$ &    	0.10~$\dagger$ 		&    0.83~$\pm$~0.20 	&    	0.10~$\dagger$ 		&    0.77\\ 
5370 $^{ }$ &    	0.10~$\dagger$ 		&    0.84~$\pm$~0.09 	&    	0.10~$\dagger$		&    0.76\\ [1ex] 
\hline 

\tablecomments{
Only epochs marked with an asterisk are used in the subsequent analysis,
as they provide the most reliable constraints on the refitted optical depth. Values
marked with $\dagger$ denote instances where the parameter pegged
at the reasonable imposed upper or lower limits that we set in the
fitting routine; the corresponding parameter ($\tau_0$ or spectral
index $\alpha$) was therefore evaluated at this fixed value and its
errors should be regarded with caution.
}

\end{tabular} 
\end{table*}

\subsection{Hydrodynamical Simulation}\label{hydro}

We based our work on the 1-dimensional hydrodynamical
simulations by \citet{Dwarkadas2010} which provides the hydrodynamic
and kinetic information of the ejecta-CSM interaction.  Without detailed
calculations and an examination of the microscopic physics, which 
is beyond the scope of this paper, it is impossible to calculate the
behaviour of the magnetic field and the acceleration of particles
to relativistic energies that dictate the strength of the synchrotron
emission. One must therefore appeal to simpler scaling relations
to infer the behaviour of these quantities in terms of others that
are well understood. \citet{Chevalier1996} has outlined several
possibilities for how the synchrotron luminosity parameters might
scale, although many other variations may exist. The radiating electron
energy density ($\propto K$) and the magnetic energy density
($\propto B^2$) may scale with the thermal energy density ($\propto P$),
since these are thought to be built up by high-pressure turbulent
motions in the shock, or may also scale with the thermal particle
density if a constant fraction of the thermal particle density is
injected in the acceleration process. Another possibility is that
the magnetic energy density may decrease as the surface
area ($B^2$~$\propto$~$r^{-2}$), as the magnetic field is carried out from
the surface of the progenitor via the wind. These simple physical
scalings, however, may not provide a complete picture of the likely
complex physics involved, as evidenced by the spectral index flattening
and the decoupling of the X-ray and radio light curves in SN 1987A, for
instance. Although there is a physical basis for the possible
energy density scalings, the only SN in which such scaling relations
have been successfully constrained (albeit somewhat controversially)
is SN 1993J (e.g., \citealt{Fransson1998}; see also \citealt{Amy2001}
for alternative viewpoint), where the magnetic energy
density was derived independently from SSA through very early
monitoring, and radius constraints were provided by VLBI data (e.g.,
\citealt{Marcaide1997}; \citealt{Bartel2000}), thus removing several of
the degeneracies that force the need for the scaling relations in the
first place. We discuss the results of applying a few particular
scaling models in Sec.~\ref{light_curve_sim}.

As mentioned in the previous section, SSA should play a role in the spectral
appearance of \hbox{SN\,1996cr} at very early times. To incorporate SSA into
our radio models in a self-consistent way, we first determined the unabsorbed
brightness temperature for each frequency of our radio model vs. time, assuming a
spherical shell morphology that evolves with the shock radius. We then
applied SSA to the full radio model assuming \hbox{$\tau_{\rm{SSA}}$~$\sim$~1}
at the critical frequency where the brightness temperature surpassed
\hbox{$\approx$~$3\times10^{11}$~K} \citep{Readhead1994}. Therefore, the
radio model is absorbed by this model-dependent SSA, in addition to
FFA whose constraints on $\tau_{\rm{ff}}$ are presented in Table~\ref{tab:constraints_tau}.

\section{Results}\label{results}

We initially characterise the ionised region surrounding \hbox{SN\,1996cr}
and compare the radio constraints on the CSM free electron density to the
expectations from hydrodynamic simulation. The simulated radio light
curves are then presented for a few basic scaling models, and are evaluated
by comparing their individual features to those of the observed light
curves. These results set the scope for further research on this object. 

\subsection{Empirical Constraints on the CSM}\label{CSM_constraints}

The constraints on the FFA optical depth allow for the study of the CSM
ionisation by providing limits on the density of free electrons. This in
turn can be used as a relatively direct cross-check on the simulation
density distribution, since the free electron density provides a strong lower
limit on the total mass in the CSM assuming full ionisation. For simplicity,
we considered a CSM composition of H only, and adopted a scaling
factor of 20 (based on the X-ray derived abundances and resulting mean
molecular weight of the CSM) to scale our results to the CSM composition
of H, He and metals presented for \hbox{SN\,1996cr} in \citet{Dwarkadas2010}.

\begin{figure}
\hglue-0.8cm{\includegraphics[width=9.5cm]{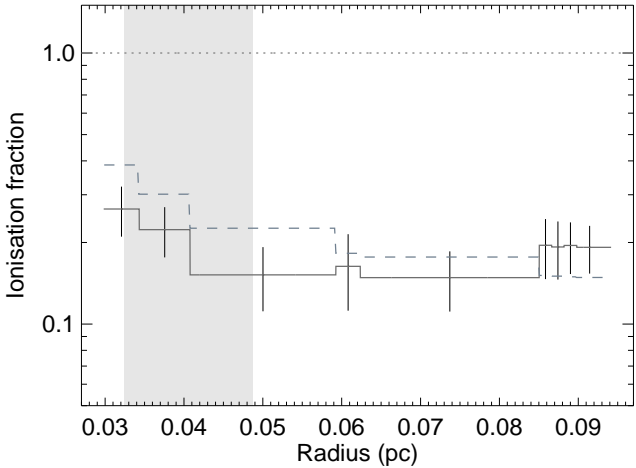}}
\caption{The CSM ionisation fraction and 1$\sigma$ error computed from
constraints on the FFA optical depth (solid line), and the theoretical
ionisation fraction for comparison (dashed line), for a CSM
temperature of $10^{4}$~K. The presumed location
of the dense shell is denoted by the gray-shaded region.}
\label{fig:ion_fraction}
\end{figure}

Each observation epoch has an approximately corresponding time step in
the simulation, and a shock radius is assigned to it based on the
position of the forward shock at that time. The free electron density
in front of the shock can be determined from the difference in the
values of the optical depth between two radio epochs
(Eq.~\ref{eq:ffa}). This requires setting a value $N_{\rm e}$ for the
outermost shock radius (latest radio epoch) from which to iterate back
through the CSM to the innermost shock radius (earliest radio epoch),
which we assumed to be $N_{\rm e}$~=~10$^4$~cm$^{-3}$, corresponding
to an average value found in H{\sc ii} regions. We adopt this value
since \hbox{SN\,1996cr} appears to reside at the centre of a diffuse
H$\alpha$ clump 5--10~pc in diameter (cf. figs. 3 and 5 of
\citealt{Bauer2008}), which is presumably a large extended H{\sc ii}
region ionised by several massive stars. For simplicity, we assumed
that the CSM has a uniform temperature, and investigated a range of
initial CSM temperatures between $T$~=~10$^{4}$--10$^{6}$~K, since
these were not directly constrained by the hydrodynamical
simulation. This range should be appropriate for a wide range of RSG
wind densities (e.g., \citealt{Fransson1998} and references therein).
Such temperature variations will lead to proportionally higher
ionisation fractions, although the radial form appears to remain
relatively steady. The shock radii and constraints on $N_{\rm e}$ are
listed in Table~\ref{tab:constraints} for $T$~=~10$^{4}$~K. The number
densities of total electrons that are available to be ionised are
obtained from the simulation over the same CSM regions delimited by
the radio epochs. Finally, the CSM ionsation fraction is calculated
directly from the free and total electron densities, and is shown in
Fig.~\ref{fig:ion_fraction}.

\begin{table}
\caption{Selected epochs with corresponding shock radii and constraints
on $N_{\rm e}$, determined from the hydrodynamic
simulation and constraints on the FFA optical depth.\label{tab:constraints}} 
\centering 
\begin{tabular}{c c c} 
\hline\hline Days since explosion & Radius ($10^{17}$ cm) & $N_{\rm e}$ ($10^{4}$~cm$^{-3}$)\\ [0.5ex]	
\hline 
 530 &   0.918 &  (2.59~$\pm$~0.16)\\
 897 &   1.058 & (2.14~$\pm$~0.14)\\
 1607 &  1.256 & (1.32~$\pm$~0.23)\\
 3039 &  1.825 & (1.06~$\pm$~0.26)\\ 
 3270 &  1.921 & (0.94~$\pm$~0.14)\\
 4737 &  2.620 & (1.05~$\pm$~0.16)\\
 4849 &  2.667 & (1.03~$\pm$~0.14)\\
 4923 &  2.717 & (1.03~$\pm$~0.08)\\
 4959 &  2.766 & 1.00 \\ [1ex] 
\hline 
\end{tabular} 
\end{table}

The radio-constrained free electron densities appear fully in line
with (i.e., lower than) the total electron densities from the
hydrodynamical simulation for temperatures lower
than 10$^{5}$~K. Overall, the CSM ionisation fraction
decreases slightly with radius; this is fully consistent with our
expectations, based on recombination after a strong
flash ionisation from the SN outburst ionising
our adopted CSM density distribution. We are relatively confident
in our adopted value of $N_{\rm e}$ for the latest radio epoch, as
all values lower than $5\times10^{4}$~cm$^{-3}$ appear stable,
yielding ionisation fractions that are consistent with the simulation,
while values over this tend to drive the ionisation in all shells
to significantly higher ionisation fractions.

As an additional check, we obtain the CSM neutral fraction
directly from the ionised fraction, and use it to calculate
the neutral hydrogen column, which we then compare to the
one obtained from spectral fits to the X-ray data. Between 1999 June
($\sim$1450 d) and 2009 January ($\sim$5000 d), the CSM neutral
fraction implies a neutral hydrogen column density of
\hbox{N$_{\rm H}$ = (9.3~$\pm$~2.2)~$\times$~$10^{20}$~amu~cm$^{-2}$}, which
is roughly consistent with a physical measurement of
\hbox{N$_{\rm H}$ = 1.8~$\times$~$10^{21}$~amu~cm$^{-2}$}
presented in \citet{Dwarkadas2010}, based on differential spectral fits to
the 2000 and 2009 HETG data. Although there may be a slight difference
in the regions being considered for both these column densities, it is
perhaps more likely that these values differ as a result of the latter
being constrained by model fits to the X-ray spectrum, which can be
easily affected by low-energy spectral lines.

\begin{figure*}
\hglue-0.7cm{\includegraphics[width=18cm]{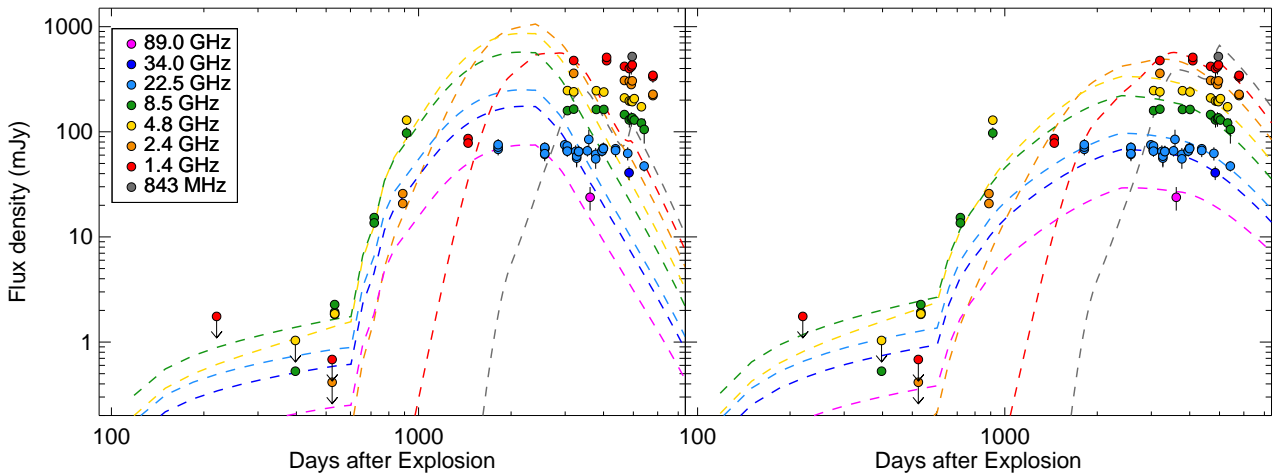}}
\caption{Simulated light curves (lines) and observed radio flux densities (circles)
for \hbox{SN\,1996cr}. Left: The `standard' model scales $K$ and $B^2$ as
the thermal energy density (erg~cm$^{-3}$). We fixed the constants of
proportionality for the $K$ and $B^2$ scalings at a radius
of 10$^{16}$~cm to be $\approx$3.4~$\times$~10$^{-5}$ and $\approx$1.3
respectively, so as to best fit the data. This model provides an
intrinsic flux rise at $\sim$700~d but fails to match the late-time plateau
region. Right: $K$ scales as the thermal particle density (cm$^{-3}$) and
$B^2$ scales as the surface area (cm$^{-2}$). The constants of
proportionality for $K$ and $B^2$ at a radius of 10$^{16}$~cm are fixed
to $\approx$4.2~$\times$~10$^{17}$ and $\sim$10$^{26}$
respectively. Although the simulated light curve provide a better
match to the late-time data, the flux at day $\sim$900 is nearly an
order of magnitude lower than observations.}
\label{fig:light_curves_1}
\end{figure*}

We can also crudely estimate the evolution of the CSM ionisation
following shock breakout on theoretical grounds to obtain an
interesting consistency check with the hydrodynamic simulation. A
SN explosion with an average output energy of $\sim$10$^{51}$~erg
will radiate (0.5--2.3)~$\times$~10$^{57}$~ionising photons during the first
1.5 days \citep{Ensman1992}, which at the higher end is sufficient to 
ionise up to $\sim$2M$_\odot$ of material \citep{Lundqvist1996}, assuming the
CSM is pure H. The CSM mass in the hydrodynamic simulation is about 2.1M$_\odot$, and thus
could possibly be almost completely ionised by the shock breakout. We
therefore assume that the entire model CSM is initially ionised, which
provides an upper bound on the true ionisation of the CSM; the true CSM is
likely to be less ionised if it is not distributed in a spherically
symmetric manner, has a mass significantly higher than 2M$_\odot$, or the explosion
energy is lower. We model the recombination using the initial CSM density
distribution from the hydrodynamical simulation (for which we assume
a CSM composition of pure H, and scale to a composition of H, He
and metals with our assumed scaling factor) and the H{\sc ii} recombination
rates given in \citet[i.e., $\alpha_B$~=~$2.6\times10^{-13}$~cm$^{3}$s$^{-1}$ for
our assumed CSM temperature]{Seager2000}. As two examples, we note
that the CSM densities of free electrons at 0.04~pc (within the shell) and
0.12~pc (within the assumed RSG wind) are 1.5$\times$10$^{6}$ and
6$\times$10$^{4}$~cm$^{-3}$, and will completely recombine in 0.08
and 2.1~yrs, respectively.

The X-ray radiation from the
SN shock provides an ongoing source of energy for reionisation
of recently recombined material. To model the amount
of ionising photons that the shock injects into the CSM as a function of
time, we assume that the ionising photon rate scales linearly with the
X-ray light curve. Using the same absorbed non-equilibrium ionisation
model adopted in \citet{Dwarkadas2010}, we normalised the photon
rate to 2.7~$\times$~$10^{48}$~\hbox{photons~s$^{-1}$} on day $\sim$1670
to match the absorbed photon rate from the 2001 August XMM spectrum. Our
simulation shows that as the shock moves through the CSM, it largely reionises the
immediate region around it (i.e., the bin it is in), but nothing more. Thus, the vast
majority of the line-of-sight ionisation we observe comes from the initial
shock-breakout ionisation. The resulting CSM ionisation fraction is shown for comparison in
Fig.~\ref{fig:ion_fraction}. From this model, the theoretical neutral
column density in the $~$1450--5000~d period is
\hbox{N$_{\rm H}$ = (9.4~$\pm$~1.9)~$\times$ $10^{20}$~amu~cm$^{-2}$}, in
accord with the value from the HETG 2000--2009 data. 

\subsection{Simulation of the Radio Light Curve}\label{light_curve_sim}

As a test of our methods, we initially computed radio light curves for a
hydrodynamic simulation of a self-similar shock expansion to demonstrate
that we could correctly reproduce the decline rates of the synchrotron luminosity
presented in \citet{Chevalier1996} for the different models therein. We then
applied this code to the more complex simulation of \citet{Dwarkadas2010},
scaling the simulated light curves to the radio data.

For \hbox{SN\,1996cr} there are three conditions that the observations require
if the model is to successfully simulate the light curve. (1) Low emission while the
ejecta is moving within the sparse cavity, since we have upper limits for this from
both X-rays and radio. (2) A sharp increase in flux from $\sim$700 to $\sim$1000~d,
as the SN ejecta initially impacts and interacts with the dense shell. The fluxes
in the 4.8 and 8.5~GHz bands rise together, which indicates that the rise is intrinsic
rather than the result of diminishing absorption, since FFA (or SSA for that matter)
will affect the bands differently. (3) A flattening of the light curve between days $\sim$1000
and $\sim$4000, with the beginning of a decrease/turnover at very late times.

Fig.~\ref{fig:light_curves_1} shows the simulated radio light curves for
two different models. The standard model in \citet{Chevalier1996}, relating
$K$ and $B^2$ to the thermal energy density, proved inadequate in our
simulations. It provides a fair match to the luminosity rise around day
700, but the subsequent peak in luminosity decreases too quickly as
the forward shock leaves the shell. The second model, which relates
$K$ to the thermal particle density and $B^2$ to the surface area,
does a better job at fitting the late-time data, but fails however to fit
the data from the rise, since the simulated flux around day 900 is almost
one order of magnitude lower than the observations. In our model we
have assumed that the radio emission is generated from both the forward
and reverse shocks, as the forward shock-only light curve resulted in
poor fits to the data (i.e., lack of plateau region followed by a sharp falloff).

More complex scalings may allow us to potentially fit the entire radio
light curve, but remain ad hoc due to our currently poor understanding of
the detailed microphysics that ultimately drives the large-scale radio
emission. This obviously makes it difficult to assess whether the deviations
between radio data and models stem from our simplistic approach,
which assumes that the emission at all times and all locations arises
from both shocks, or alternatively, may still be limited by the hydrodynamic
simulation. While the simulation has some difficulty reproducing a similar
feature in the X-ray light curve \citep{Dwarkadas2010}, it is able to reproduce
the X-ray spectra and light curves, which are more directly and fundamentally
tied to physical parameters (e.g., density, emissivity, and ionisation equilibrium),
and even allow to separate X-ray emission from the forward shock, which
dominates during the first seven years, and reverse shock, which dominates
thereafter. Our radio calculations, on the other hand, do not allow this level
of accuracy or differentiation.

\section{Discussion $\&$ Conclusions}\label{discussion}

We have presented an up-to-date sequence of radio light curves of
\hbox{SN\,1996cr}. Using an analytic model that describes the radio emission
from SNe, combined with a hydrodynamic simulation of the synchrotrom-emitting
regions, we simulated the observed radio light curves of \hbox{SN\,1996cr} 
to further constrain the density of the CSM, particularly in the sparse region interior
to the shell, for which there are no firm X-ray constraints implemented in the
hydrodynamic simulation.

To this end, we explored some basic scalings that relate the hydrodynamics
of the shock to the magnetic field and the radiating electron distribution
that are required for synchrotron emission. For all of the models studied,
we found that the behaviour in certain periods of our simulated radio light
curves matched the functional form of our observations relatively well, but
invariably failed to match the most tightly constraining temporal features,
typically the rise around $\sim$700 d and/or the plateau region between
$\sim$2000--5000 d.

The simulation shown on the right panel in Fig.~\ref{fig:light_curves_1} (with 
$K$ $\propto$ thermal particle density and $B^2$ $\propto$ surface area) matches
the radio light curves reasonably well from a radius of $\approx$0.04~pc
outwards. The different scalings that we studied have their share of assumptions and
potential failings. However, the mismatch interior to $\approx$0.04~pc coincides with
the lack of strong constraints on the simulation parameters from the X-ray
light curve at similar radii, which may mean that the inner region of the
dense shell is not being adequately modelled. To improve the early
discrepancies, we need to understand how the magnetic field and accelerated
particle distribution actually scale as the SNe shock collides with a dense CSM
shell, and/or have the simulation potentially consider (1) a more concentrated
shell mass along the inner edge, (2) clumpy structure that could provide
modifications to both the ionisation and radio flux, (3) asymmetric mass
distribution to spread out/flatten the peak, or (4) modification to the SN
parameters (e.g., ejecta density slope, the ejecta mass and explosion
energy). Changes to the simulations, however, cannot be too large, since
the non-equilibrium ionisation model spectra already match the HETG X-ray
spectra well over several epochs.

Geometrical considerations, such as possible asymmetries in the wind, in the
SN explosion or in the dense shell, were not included in our calculations. \citet{Dewey2011}
recently showed that \hbox{SN\,1996cr} might in fact not be spherically
symmetric. Nevertheless, our assumption that a 1D hydrodynamic simulation
is appropriate to simulate the radio light curve is at least justified
by the fact that it also reproduces the broad features of the
observed X-ray light curve \citep{Dwarkadas2010}.

Of course, the X-ray emission is thermal, while the radio emission is
non-thermal synchrotron. These are very different mechanisms,
arising from potentially distinct regions and scaling with different
physical parameters; the microphysics and mechanics of which
are poorly understood or constrained in the radio. Thus we should not
be surprised to see a decoupling of the simulated X-ray and radio
light curves. Clearly, the physics that generates the X-ray emission
is much simpler than that of the radio, as it relates only to the
density, emissivity, and ionisation (in)equilibrium of the emitting
material. This close link between the hydrodynamical parameters of
the shock and a correct simulation of the observed X-ray light curve
lends more support to the hydrodynamic simulation as accurately
descriptive of the dynamics of the shock. We attempted to simulate the
radio light curve using simple assumptions for the magnetic field
and the electron distribution, which do not appear to be appropriate
for complex objects such as \hbox{SN\,1996cr}. The use of simple
scaling relations, combined with hydrodynamic simulations, also failed
to adequately reproduce the sharp rise and subsequent linear increase
of radio emission in \hbox{SN\,1987A} \citep{Dwarkadas2007b,Dwarkadas2007c}. These
are precautions to be considered when trying to constrain the
circumstellar density of these objects using simple radio models,
because even if an X-ray-based density model is available, it may
not be able to accurately model the radio light curve while avoiding
the degeneracies that arise in trying to determine viable density scalings. 

SN 1996cr remains quite bright at nearly all wavelengths, and future radio and
X-ray monitoring will be able to constrain the apparent decline, which should
place stronger constraints on the outer wind and CSM density distribution. Future
monitoring with the LBA and ALMA will be able to resolve the radio
morphology of \hbox{SN\,1996cr} in greater detail to constrain its spatial
expansion (i.e., current velocity) and emission regions.

\section*{Acknowledgments}

CM and FEB would like to thank Andr\'{e}s Jord\'{a}n for
suggestions. Special thanks to Juergen Ott for providing the 89~GHz
reduced data and Shari Breen for early access to her data. We
acknowledge support from CONICYT-Chile under grants FONDECYT 1101024
(CM, FEB), ALMA-CONICYT 31100004 (FEB), and FONDAP-CATA 15010003
(FEB), Basal-CATA grant PFB-06/2007 (FEB), the Iniciativa
Cient\'{\i}fica Milenio through the Millennium Center for Supernova
Science grant P10-064-F with input from "Fondo de Innovaci\'{o}n para
la Competitividad, del Ministerio de Econom\'{\i}a, Fomento y Turismo
de Chile`` (FEB), and Chandra X-ray Center grants SAO GO9-0086,
GO0-11095, and GO1-12095B (FEB, VVD).

FEB thanks the ATNF Distinguished Visitor program. The ATCA is
part of the Australia Telescope National Facility which is funded by
the Commonwealth of Australia for operation as a National Facility
managed by CSIRO. This paper includes archived data obtained
through the Australia Telescope Online Archive
(http://atoa.atnf.csiro.au). The MOST is operated by the University of Sydney
with support from the Australian Research Council.

\label{lastpage}

\section{Appendix}
\appendix

Below we provide flux density tables for the calibrators and nucleus
measurements (Tables~\ref{tab:data_radio_cal} and
\ref{tab:data_radio_nuc} respectively) which are shown 
in the lower panels of Fig.~\ref{fig:data_radio}. These were
reduced in the same manner as the flux densities of
\hbox{SN\,1996cr}, and demonstrate that the variability we see is
real. The full tables are available in the online journal.

\renewcommand*\thetable{A\arabic{table}}

\begin{landscape}

\def\arraystretch{1.3}

\begin{deluxetable}{lcc|cc|cc|cc|cc|cc|cc|ccl}


\tabletypesize{\scriptsize}
\tablewidth{0pt}
\tablecaption{Calibrators\label{tab:data_radio_cal}}
\tablehead{
\colhead{Obs. Date} &
\colhead{Obs. ID} &
\colhead{Calibrator} &
\multicolumn{2}{|c|}{W-Band} &
\multicolumn{2}{|c|}{K$_a$-Band} &
\multicolumn{2}{|c|}{K-Band} &
\multicolumn{2}{|c|}{X-Band} &
\multicolumn{2}{|c|}{C-Band} &
\multicolumn{2}{|c|}{S-Band} &
\multicolumn{2}{|c}{L-Band} \\
& 
& 
&
\multicolumn{1}{|c}{$\nu$} &
\multicolumn{1}{c|}{$S_{\nu}$} &
\multicolumn{1}{|c}{$\nu$} &
\multicolumn{1}{c|}{$S_{\nu}$} &
\multicolumn{1}{|c}{$\nu$} &
\multicolumn{1}{c|}{$S_{\nu}$} &
\multicolumn{1}{|c}{$\nu$} &
\multicolumn{1}{c|}{$S_{\nu}$} &
\multicolumn{1}{|c}{$\nu$} &
\multicolumn{1}{c|}{$S_{\nu}$} &
\multicolumn{1}{|c}{$\nu$} &
\multicolumn{1}{c|}{$S_{\nu}$} &
\multicolumn{1}{|c}{$\nu$} &
\multicolumn{1}{c}{$S_{\nu}$}\\
\colhead{(1)}&
\colhead{(2)}&
\colhead{(3)}&
\colhead{(4)}&
\colhead{(5)}&
\colhead{(6)}&
\colhead{(7)}&
\colhead{(8)}&
\colhead{(9)}&
\colhead{(10)}&
\colhead{(11)}&
\colhead{(12)}&
\colhead{(13)}&
\colhead{(14)}&
\colhead{(15)}&
\colhead{(16)}&
\colhead{(17)}

}

\startdata

1995-03-03 & C204a & 1329-665 &     ---     &    ---     &    ---       &    ---     &     ---       &    ---    &         8640 &  348.0 $\pm$  11.6 &         4800 &  854.0 $\pm$  14.1 &     ---     &    ---     &     ---     &    ---      \\
1995-03-03 & C204a & 1329-665 &     ---     &    ---     &    ---       &    ---     &     ---       &    ---    &         8640 &  348.0 $\pm$  11.6 &         4800 &  854.0 $\pm$  14.1 &     ---     &    ---     &     ---     &    ---      \\
1995-03-30 & C363 & 1329-665 &     ---     &    ---     &    ---       &    ---     &     ---       &    ---    &     ---      &    ---     &     ---     &    ---     &     ---     &    ---     &         1418 & 2680.0 $\pm$  34.0  \\
1995-04-16 & C418 & 1329-665 &     ---     &    ---     &    ---       &    ---     &     ---       &    ---    &     ---      &    ---     &     ---     &    ---     &         2378 & 1660.0 $\pm$  17.1 &         1418 & 2650.0 $\pm$  50.4  \\
1995-05-27 & C204a & 1329-665 &     ---     &    ---     &    ---       &    ---     &     ---       &    ---    &         8640 &  347.0 $\pm$  27.7 &         4800 &  855.0 $\pm$  15.8 &     ---     &    ---     &     ---     &    ---      \\
1995-05-27 & C204a & 1329-665 &     ---     &    ---     &    ---       &    ---     &     ---       &    ---    &         8640 &  347.0 $\pm$  27.7 &         4800 &  855.0 $\pm$  15.8 &     ---     &    ---     &     ---     &    ---      \\
1995-06-07 & C204a & 1329-665 &     ---     &    ---     &    ---       &    ---     &     ---       &    ---    &         8640 &  348.0 $\pm$  12.9 &         4800 &  853.0 $\pm$  15.3 &     ---     &    ---     &     ---     &    ---      \\
1995-06-07 & C204a & 1329-665 &     ---     &    ---     &    ---       &    ---     &     ---       &    ---    &         8640 &  348.0 $\pm$  12.9 &         4800 &  853.0 $\pm$  15.3 &     ---     &    ---     &     ---     &    ---      \\
1995-07-27 & C204c & 1329-665 &     ---     &    ---     &    ---       &    ---     &     ---       &    ---    &     ---      &    ---     &     ---     &    ---     &         2368 & 1680.0 $\pm$  15.1 &         1376 & 2860.0 $\pm$  19.9  \\
1995-08-03 & C363 & 1329-665 &     ---     &    ---     &    ---       &    ---     &     ---       &    ---    &     ---      &    ---     &     ---     &    ---     &     ---     &    ---     &         1418 & 2480.0 $\pm$  82.3  \\
1996-02-06 & CT09 & 1718-649 &     ---     &    ---     &    ---       &    ---     &     ---       &    ---    &     ---      &    ---     &     ---     &    ---     &     ---     &    ---     &         1664 & 3640.0 $\pm$  61.4  \\
1996-08-01 & C466 & 1236-684 &     ---     &    ---     &    ---       &    ---     &     ---       &    ---    &         8640 &  355.0 $\pm$  15.6 &         4800 &  283.0 $\pm$   9.6 &     ---     &    ---     &     ---     &    ---      \\
1996-12-05 & C505 & 1236-684 &     ---     &    ---     &    ---       &    ---     &     ---       &    ---    &     ---      &    ---     &     ---     &    ---     &         2368 &  572.0 $\pm$  13.3 &         1384 &  747.0 $\pm$  15.5  \\
1996-12-15 & C586 & 1549-790 &     ---     &    ---     &    ---       &    ---     &     ---       &    ---    &         8512 & 2320.0 $\pm$  18.0 &         4928 & 3490.0 $\pm$  16.7 &     ---     &    ---     &     ---     &    ---      \\
1996-12-15 & C586 & 1549-790 &     ---     &    ---     &    ---       &    ---     &     ---       &    ---    &         8512 & 2320.0 $\pm$  18.0 &         4928 & 3490.0 $\pm$  16.7 &     ---     &    ---     &     ---     &    ---      \\
1997-06-17 & V100C & 1045-62 &     ---     &    ---     &    ---       &    ---     &     ---       &    ---    &         8425 & 1640.0 $\pm$  59.8 &     ---     &    ---     &     ---     &    ---     &     ---     &    ---      \\
1997-06-17 & V100C & 1045-62 &     ---     &    ---     &    ---       &    ---     &     ---       &    ---    &         8425 & 1430.0 $\pm$  52.7 &     ---     &    ---     &     ---     &    ---     &     ---     &    ---      \\
1997-12-05 & V099 & 1104-445 &     ---     &    ---     &    ---       &    ---     &     ---       &    ---    &     ---      &    ---     &     ---     &    ---     &         2268 & 2890.0 $\pm$  79.8 &     ---     &    ---      \\
1997-12-05 & V099 & 1104-445 &     ---     &    ---     &    ---       &    ---     &     ---       &    ---    &     ---      &    ---     &     ---     &    ---     &         2268 & 2610.0 $\pm$  32.2 &     ---     &    ---      \\
1997-12-31 & C694 & 1414-59 &     ---     &    ---     &    ---       &    ---     &     ---       &    ---    &         6024 &  819.0 $\pm$  38.5 &         4800 &  891.0 $\pm$  20.2 &     ---     &    ---     &     ---     &    ---      \\
1999-06-19 & C788 & 1329-665 &     ---     &    ---     &    ---       &    ---     &     ---       &    ---    &     ---      &    ---     &     ---     &    ---     &     ---     &    ---     &         1418 & 2600.0 $\pm$  71.2  \\
1999-06-19 & C788 & 1329-665 &     ---     &    ---     &    ---       &    ---     &     ---       &    ---    &     ---      &    ---     &     ---     &    ---     &     ---     &    ---     &         1384 & 2670.0 $\pm$  14.0  \\
2000-06-19 & V137 & 1424-418 &     ---     &    ---     &    ---       &    ---     &        22215 & 1650.0 $\pm$  35.7 &     ---      &    ---     &     ---     &    ---     &     ---     &    ---     &     ---     &    ---      \\
2000-06-19 & V137 & 1424-418 &     ---     &    ---     &    ---       &    ---     &        22203 & 1580.0 $\pm$  56.8 &     ---      &    ---     &     ---     &    ---     &     ---     &    ---     &     ---     &    ---      \\
2000-06-19 & V137 & 1424-418 &     ---     &    ---     &    ---       &    ---     &        22190 & 1590.0 $\pm$  35.1 &     ---      &    ---     &     ---     &    ---     &     ---     &    ---     &     ---     &    ---      \\
2002-07-18 & V137 & 1610-771 &     ---     &    ---     &    ---       &    ---     &        22224 & 2100.0 $\pm$  21.0 &     ---      &    ---     &     ---     &    ---     &     ---     &    ---     &     ---     &    ---      \\
2002-07-18 & V137 & 1610-771 &     ---     &    ---     &    ---       &    ---     &        22216 & 2100.0 $\pm$  20.4 &     ---      &    ---     &     ---     &    ---     &     ---     &    ---     &     ---     &    ---      \\
2002-07-18 & V137 & 1610-771 &     ---     &    ---     &    ---       &    ---     &        22201 & 2070.0 $\pm$  36.5 &     ---      &    ---     &     ---     &    ---     &     ---     &    ---     &     ---     &    ---      \\
2002-07-18 & V137 & 1610-771 &     ---     &    ---     &    ---       &    ---     &        22187 & 2100.0 $\pm$  11.1 &     ---      &    ---     &     ---     &    ---     &     ---     &    ---     &     ---     &    ---      \\
2002-07-18 & V137 & 1610-771 &     ---     &    ---     &    ---       &    ---     &        22172 & 2100.0 $\pm$  19.4 &     ---      &    ---     &     ---     &    ---     &     ---     &    ---     &     ---     &    ---      \\
2003-09-07 & C1224 & 1414-59 &     ---     &    ---     &    ---       &    ---     &        23659 & 1190.0 $\pm$  17.8 &     ---      &    ---     &     ---     &    ---     &     ---     &    ---     &     ---     &    ---      \\
2003-11-03 & C1049 & 1414-59 &     ---     &    ---     &    ---       &    ---     &        19008 & 1150.0 $\pm$  30.7 &     ---      &    ---     &     ---     &    ---     &     ---     &    ---     &     ---     &    ---      \\
2003-11-03 & C1049 & 1414-59 &     ---     &    ---     &    ---       &    ---     &        16960 & 1160.0 $\pm$  30.4 &     ---      &    ---     &     ---     &    ---     &     ---     &    ---     &     ---     &    ---      \\
2003-11-04 & C1049 & 1414-59 &     ---     &    ---     &    ---       &    ---     &        22796 & 1200.0 $\pm$  69.7 &     ---      &    ---     &     ---     &    ---     &     ---     &    ---     &     ---     &    ---      \\
2003-11-04 & C1049 & 1414-59 &     ---     &    ---     &    ---       &    ---     &        21056 & 1220.0 $\pm$  59.9 &     ---      &    ---     &     ---     &    ---     &     ---     &    ---     &     ---     &    ---      \\
2003-11-08 & C1049 & 1414-59 &     ---     &    ---     &    ---       &    ---     &     ---       &    ---    &         8256 & 1230.0 $\pm$  33.0 &         5056 &  954.0 $\pm$  24.1 &     ---     &    ---     &     ---     &    ---      \\
2004-04-02 & C1424 & 1338-58 &     ---     &    ---     &    ---       &    ---     &     ---       &    ---    &         8640 &  433.0 $\pm$  18.0 &         4800 &  918.0 $\pm$  14.6 &         2368 & 1880.0 $\pm$  27.6 &         1384 & 2800.0 $\pm$  35.3  \\
2004-05-20 & CX065 & 1414-59 &     ---     &    ---     &    ---       &    ---     &        22087 & 1100.0 $\pm$  50.0 &     ---      &    ---     &     ---     &    ---     &     ---     &    ---     &     ---     &    ---      \\
2004-06-13 & V176 & 1414-59 &     ---     &    ---     &    ---       &    ---     &        22209 & 1140.0 $\pm$  38.0 &     ---      &    ---     &     ---     &    ---     &     ---     &    ---     &     ---     &    ---      \\
2004-06-13 & V176 & 1414-59 &     ---     &    ---     &    ---       &    ---     &        22195 & 1150.0 $\pm$  38.5 &     ---      &    ---     &     ---     &    ---     &     ---     &    ---     &     ---     &    ---      \\
2004-08-01 & V176B & 1414-59 &     ---     &    ---     &    ---       &    ---     &        22209 &  993.0 $\pm$  28.8 &     ---      &    ---     &     ---     &    ---     &     ---     &    ---     &     ---     &    ---      \\
2004-08-01 & V176B & 1414-59 &     ---     &    ---     &    ---       &    ---     &        22195 &  993.0 $\pm$  31.0 &     ---      &    ---     &     ---     &    ---     &     ---     &    ---     &     ---     &    ---      \\
2005-03-14 & C1368 & 1414-59 &     ---     &    ---     &    ---       &    ---     &        22193 &  877.0 $\pm$  20.9 &     ---      &    ---     &     ---     &    ---     &     ---     &    ---     &     ---     &    ---      \\
2005-04-20 & C1368 & 1414-59 &     ---     &    ---     &    ---       &    ---     &        22194 &   85.6 $\pm$  20.0 &     ---      &    ---     &     ---     &    ---     &     ---     &    ---     &     ---     &    ---      \\
2005-10-27 & C1049 & 1414-59 &     ---     &    ---     &    ---       &    ---     &        18752 &  941.0 $\pm$  45.5 &     ---      &    ---     &     ---     &    ---     &     ---     &    ---     &     ---     &    ---      \\
2005-10-27 & C1049 & 1414-59 &     ---     &    ---     &    ---       &    ---     &        21056 &  898.0 $\pm$  66.0 &     ---      &    ---     &     ---     &    ---     &     ---     &    ---     &     ---     &    ---      \\
2005-11-13 & C1049 & 1414-59 &     ---     &    ---     &    ---       &    ---     &     ---       &    ---    &         8640 &  988.0 $\pm$  30.0 &         4800 &  966.0 $\pm$  25.1 &     ---     &    ---     &     ---     &    ---      \\
2006-04-29 & C1049 & 1414-59 &     ---     &    ---     &    ---       &    ---     &        18752 &  806.0 $\pm$  66.3 &     ---      &    ---     &     ---     &    ---     &     ---     &    ---     &     ---     &    ---      \\
2006-04-29 & C1049 & 1414-59 &     ---     &    ---     &    ---       &    ---     &        21056 &  804.0 $\pm$  93.1 &     ---      &    ---     &     ---     &    ---     &     ---     &    ---     &     ---     &    ---      \\
2006-06-08 & CX114 & 1352-63 &     ---     &    ---     &    ---       &    ---     &        22193 &  817.0 $\pm$  23.0 &     ---      &    ---     &     ---     &    ---     &     ---     &    ---     &     ---     &    ---      \\
2006-06-08 & CX114 & 1352-63 &     ---     &    ---     &    ---       &    ---     &        20160 &  866.0 $\pm$  32.8 &     ---      &    ---     &     ---     &    ---     &     ---     &    ---     &     ---     &    ---      \\
2006-06-21 & C1049 & 1414-59 &     ---     &    ---     &    ---       &    ---     &     ---       &    ---    &         8640 &  878.0 $\pm$  16.3 &         4800 &  879.0 $\pm$  13.5 &     ---     &    ---     &     ---     &    ---      \\
2006-09-14 & C1341 & 1329-665 &     ---     &    ---     &    ---       &    ---     &     ---       &    ---    &     ---      &    ---     &     ---     &    ---     &     ---     &    ---     &         1384 & 3040.0 $\pm$  90.4  \\
2006-09-14 & C1341 & 1329-665 &     ---     &    ---     &    ---       &    ---     &     ---       &    ---    &     ---      &    ---     &     ---     &    ---     &     ---     &    ---     &         1418 & 2780.0 $\pm$  90.4  \\
2007-03-22 & VT10C3 & 1352-63 &     ---     &    ---     &    ---       &    ---     &     ---       &    ---    &         8425 & 1120.0 $\pm$  50.0 &     ---     &    ---     &     ---     &    ---     &     ---     &    ---      \\
2007-06-24 & VX013A & 1352-63 &     ---     &    ---     &    ---       &    ---     &        22316 & 1060.0 $\pm$  22.1 &     ---      &    ---     &     ---     &    ---     &     ---     &    ---     &     ---     &    ---      \\
2007-06-24 & VX013A & 1352-63 &     ---     &    ---     &    ---       &    ---     &        22348 & 1190.0 $\pm$  25.5 &     ---      &    ---     &     ---     &    ---     &     ---     &    ---     &     ---     &    ---      \\
2008-04-21 & C1803 & 1352-63 &     ---     &    ---     &    ---       &    ---     &     ---       &    ---    &         8640 & 1410.0 $\pm$  33.0 &         4800 & 1530.0 $\pm$  32.9 &         2368 & 1440.0 $\pm$  41.4 &         1384 & 1250.0 $\pm$  41.7  \\
2008-08-16 & C1750 & 1352-63 &     ---     &    ---     &    ---       &    ---     &        22192 &  951.0 $\pm$  25.2 &     ---      &    ---     &     ---     &    ---     &     ---     &    ---     &     ---     &    ---      \\
2008-10-02 & C184 & 1352-63 &     ---     &    ---     &        34496 &  893.0 $\pm$  54.2 &     ---       &    ---    &     ---      &    ---     &     ---     &    ---     &     ---     &    ---     &     ---     &    ---      \\
2008-10-12 & C184 & 1352-63 &     ---     &    ---     &    ---       &    ---     &     ---       &    ---    &         8640 & 1410.0 $\pm$  31.8 &         4786 & 1610.0 $\pm$  29.0 &     ---     &    ---     &     ---     &    ---      \\
2008-10-12 & C184 & 1329-665 &     ---     &    ---     &    ---       &    ---     &     ---       &    ---    &     ---      &    ---     &     ---     &    ---     &         2496 & 1520.0 $\pm$  59.7 &         1376 & 2550.0 $\pm$  77.3  \\
2008-12-22 & C184 & 1352-63 &     ---     &    ---     &    ---       &    ---     &     ---       &    ---    &         8640 & 1280.0 $\pm$  26.4 &         4786 & 1550.0 $\pm$  30.4 &         2496 & 1400.0 $\pm$  30.4 &         1376 & 1270.0 $\pm$  25.4  \\
2009-01-28 & C184 & 1352-63 &     ---     &    ---     &    ---       &    ---     &     ---       &    ---    &         8640 & 1330.0 $\pm$  28.7 &         4786 & 1510.0 $\pm$  30.8 &         2496 & 1420.0 $\pm$  33.1 &         1376 & 1300.0 $\pm$  30.4  \\
2009-02-16 & C184 & 1352-63 &     ---     &    ---     &    ---       &    ---     &     ---       &    ---    &         8640 & 1210.0 $\pm$  27.2 &         4786 & 1490.0 $\pm$  25.4 &         2496 & 1420.0 $\pm$  39.6 &         1376 & 1250.0 $\pm$  41.4  \\
2009-04-13 & CX172 & 1352-63 &     ---     &    ---     &    ---       &    ---     &     ---       &    ---    &         8800 & 1350.0 $\pm$  18.3 &         5500 & 1560.0 $\pm$  17.3 &     ---     &    ---     &     ---     &    ---      \\
2010-01-16 & C1726 & 1352-63 &     ---     &    ---     &    ---       &    ---     &     ---       &    ---    &         9000 & 1370.0 $\pm$  21.2 &         5500 & 1500.0 $\pm$  14.3 &     ---     &    ---     &     ---     &    ---      \\
2010-05-05 & V253 & 1352-63 &     ---     &    ---     &    ---       &    ---     &        22000 &  803.0 $\pm$  32.3 &     ---      &    ---     &     ---     &    ---     &     ---     &    ---     &     ---     &    ---      \\
2010-05-07 & V253 & 1352-63 &     ---     &    ---     &    ---       &    ---     &     ---       &    ---    &         8425 & 1240.0 $\pm$  11.1 &     ---     &    ---     &     ---     &    ---     &     ---     &    ---      \\
2011-05-09 & C2479 & 1329-665 &     ---     &    ---     &    ---       &    ---     &     ---       &    ---    &     ---      &    ---     &     ---     &    ---     &         2100 & 1550.0 $\pm$  80.0 &         1418 & 2500.0 $\pm$  98.0  \\
2011-05-09 & C2479 & 1329-665 &     ---     &    ---     &    ---       &    ---     &     ---       &    ---    &     ---      &    ---     &     ---     &    ---     &         2100 & 1500.0 $\pm$  20.0 &         1613 & 2510.0 $\pm$  69.0  \\

\enddata
\begin{flushleft}

\tablecomments{
{\it Column 1:} Starting date of observation. 
{\it Column 2:} Observing program ID.
{\it Column 3:} Calibrator.
{\it Columns 4, 6, 8, 10, 12, 14, and 16:} Mean frequency of observed band in
units of MHz.
{\it Columns 5, 7, 9, 11, 13, 15, and 17:} Integrated flux densities in units of mJy as measured by {\sc uvfit}  in {\sc miriad}.
}
\end{flushleft}
\end{deluxetable}
\end{landscape}

\begin{landscape}

\def\arraystretch{1.3}

\begin{deluxetable}{lc|cc|cc|cc|cc|cc|cc|cc}

\tabletypesize{\scriptsize}
\tablewidth{0pt}
\tablecaption{Nucleus\label{tab:data_radio_nuc}}
\tablehead{
\colhead{Obs. Date} &
\colhead{Obs. ID} &
\multicolumn{2}{|c|}{W-Band} &
\multicolumn{2}{|c|}{K$_a$-Band} &
\multicolumn{2}{|c|}{K-Band} &
\multicolumn{2}{|c|}{X-Band} &
\multicolumn{2}{|c|}{C-Band} &
\multicolumn{2}{|c|}{S-Band} &
\multicolumn{2}{|c}{L-Band} \\
& 
& 
\multicolumn{1}{|c}{$\nu$} &
\multicolumn{1}{c|}{$S_{\nu}$} &
\multicolumn{1}{|c}{$\nu$} &
\multicolumn{1}{c|}{$S_{\nu}$} &
\multicolumn{1}{|c}{$\nu$} &
\multicolumn{1}{c|}{$S_{\nu}$} &
\multicolumn{1}{|c}{$\nu$} &
\multicolumn{1}{c|}{$S_{\nu}$} &
\multicolumn{1}{|c}{$\nu$} &
\multicolumn{1}{c|}{$S_{\nu}$} &
\multicolumn{1}{|c}{$\nu$} &
\multicolumn{1}{c|}{$S_{\nu}$} &
\multicolumn{1}{|c}{$\nu$} &
\multicolumn{1}{c}{$S_{\nu}$} \\
\colhead{(1)}&
\colhead{(2)}&
\colhead{(3)}&
\colhead{(4)}&
\colhead{(5)}&
\colhead{(6)}&
\colhead{(7)}&
\colhead{(8)}&
\colhead{(9)}&
\colhead{(10)}&
\colhead{(11)}&
\colhead{(12)}&
\colhead{(13)}&
\colhead{(14)}&
\colhead{(15)}&
\colhead{(16)}

}

\startdata
1995-01-20 & C363 &     ---     &    ---     &    ---       &    ---     &     ---       &    ---    &     ---      &    ---     &     ---     &    ---     &     ---     &    ---     &         1418 &   45.8 $\pm$   2.9  \\
1995-02-25 & C363 &     ---     &    ---     &    ---       &    ---     &     ---       &    ---    &     ---      &    ---     &     ---     &    ---     &     ---     &    ---     &         1418 &   65.0 $\pm$   3.1  \\
1995-03-03 & C204a &     ---     &    ---     &    ---       &    ---     &     ---       &    ---    &         8640 &   22.7 $\pm$   9.0 &         4800 &   44.9 $\pm$   4.3 &     ---     &    ---     &     ---     &    ---      \\
1995-03-03 & C204a &     ---     &    ---     &    ---       &    ---     &     ---       &    ---    &         8640 &   37.4 $\pm$   9.2 &         4800 &   45.5 $\pm$   4.4 &     ---     &    ---     &     ---     &    ---      \\
1995-03-30 & C363 &     ---     &    ---     &    ---       &    ---     &     ---       &    ---    &     ---      &    ---     &     ---     &    ---     &     ---     &    ---     &         1418 &   64.3 $\pm$   2.3  \\
1995-04-16 & C418 &     ---     &    ---     &    ---       &    ---     &     ---       &    ---    &     ---      &    ---     &     ---     &    ---     &         2768 &   53.4 $\pm$   1.0 &         1418 &   62.8 $\pm$   1.0  \\
1995-05-27 & C204a &     ---     &    ---     &    ---       &    ---     &     ---       &    ---    &         8640 &   41.1 $\pm$  11.0 &         4800 &   48.8 $\pm$   4.9 &     ---     &    ---     &     ---     &    ---      \\
1995-05-27 & C204a &     ---     &    ---     &    ---       &    ---     &     ---       &    ---    &         8640 &   48.0 $\pm$  12.0 &         4800 &   48.6 $\pm$   4.9 &     ---     &    ---     &     ---     &    ---      \\
1995-06-07 & C204a &     ---     &    ---     &    ---       &    ---     &     ---       &    ---    &         8640 &   48.7 $\pm$  12.0 &         4800 &   43.2 $\pm$   4.3 &     ---     &    ---     &     ---     &    ---      \\
1995-06-07 & C204a &     ---     &    ---     &    ---       &    ---     &     ---       &    ---    &         8640 &   49.1 $\pm$   1.2 &         4800 &   42.9 $\pm$   4.3 &     ---     &    ---     &     ---     &    ---      \\
1995-07-27 & C204c &     ---     &    ---     &    ---       &    ---     &     ---       &    ---    &     ---      &    ---     &     ---     &    ---     &         2368 &   44.5 $\pm$   3.5 &         1376 &   65.8 $\pm$   4.0  \\
1995-08-03 & C363 &     ---     &    ---     &    ---       &    ---     &     ---       &    ---    &     ---      &    ---     &     ---     &    ---     &     ---     &    ---     &         1418 &   52.4 $\pm$   3.0  \\
1996-02-06 & CT09 &     ---     &    ---     &    ---       &    ---     &     ---       &    ---    &     ---      &    ---     &     ---     &    ---     &     ---     &    ---     &         1664 &   56.5 $\pm$   1.0  \\
1996-08-01 & C466 &     ---     &    ---     &    ---       &    ---     &     ---       &    ---    &         8640 &  114.0 $\pm$  20.0 &         4800 &   83.3 $\pm$   1.7 &     ---     &    ---     &     ---     &    ---      \\
1996-12-05 & C505 &     ---     &    ---     &    ---       &    ---     &     ---       &    ---    &     ---      &    ---     &     ---     &    ---     &         2368 &   40.2 $\pm$   1.0 &         1384 &   50.5 $\pm$   1.0  \\
1996-12-15 & C586 &     ---     &    ---     &    ---       &    ---     &     ---       &    ---    &         8512 &   55.0 $\pm$   5.5 &         4928 &   48.1 $\pm$   4.8 &     ---     &    ---     &     ---     &    ---      \\
1996-12-15 & C586 &     ---     &    ---     &    ---       &    ---     &     ---       &    ---    &         8512 &   56.4 $\pm$   5.6 &         4928 &   48.6 $\pm$   4.9 &     ---     &    ---     &     ---     &    ---      \\
1997-06-17 & V100C &     ---     &    ---     &    ---       &    ---     &     ---       &    ---    &         8425 &   87.7 $\pm$  16.0 &     ---     &    ---     &     ---     &    ---     &     ---     &    ---      \\
1997-06-17 & V100C &     ---     &    ---     &    ---       &    ---     &     ---       &    ---    &         8425 &   76.9 $\pm$   5.0 &     ---     &    ---     &     ---     &    ---     &     ---     &    ---      \\
1997-12-05 & V099 &     ---     &    ---     &    ---       &    ---     &     ---       &    ---    &     ---      &    ---     &     ---     &    ---     &         2268 &   86.3 $\pm$   8.0 &     ---     &    ---      \\
1997-12-05 & V099 &     ---     &    ---     &    ---       &    ---     &     ---       &    ---    &     ---      &    ---     &     ---     &    ---     &         2268 &   56.4 $\pm$   6.0 &     ---     &    ---      \\
1997-12-31 & C694 &     ---     &    ---     &    ---       &    ---     &     ---       &    ---    &         6024 &   80.8 $\pm$   7.0 &         4800 &   67.5 $\pm$   4.7 &     ---     &    ---     &     ---     &    ---      \\
1999-06-19 & C788 &     ---     &    ---     &    ---       &    ---     &     ---       &    ---    &     ---      &    ---     &     ---     &    ---     &     ---     &    ---     &         1418 &   48.4 $\pm$   4.0  \\
1999-06-19 & C788 &     ---     &    ---     &    ---       &    ---     &     ---       &    ---    &     ---      &    ---     &     ---     &    ---     &     ---     &    ---     &         1384 &   45.4 $\pm$   4.0  \\
2000-06-19 & V137 &     ---     &    ---     &    ---       &    ---     &        22215 &   94.2 $\pm$   8.4 &     ---      &    ---     &     ---     &    ---     &     ---     &    ---     &     ---     &    ---      \\
2000-06-19 & V137 &     ---     &    ---     &    ---       &    ---     &        22203 &   95.2 $\pm$   7.0 &     ---      &    ---     &     ---     &    ---     &     ---     &    ---     &     ---     &    ---      \\
2000-06-19 & V137 &     ---     &    ---     &    ---       &    ---     &        22190 &  107.0 $\pm$   7.1 &     ---      &    ---     &     ---     &    ---     &     ---     &    ---     &     ---     &    ---      \\
2002-07-18 & V137 &     ---     &    ---     &    ---       &    ---     &        22224 &   79.3 $\pm$  10.0 &     ---      &    ---     &     ---     &    ---     &     ---     &    ---     &     ---     &    ---      \\
2002-07-18 & V137 &     ---     &    ---     &    ---       &    ---     &        22216 &   73.9 $\pm$   8.0 &     ---      &    ---     &     ---     &    ---     &     ---     &    ---     &     ---     &    ---      \\
2002-07-18 & V137 &     ---     &    ---     &    ---       &    ---     &        22201 &   83.2 $\pm$   8.7 &     ---      &    ---     &     ---     &    ---     &     ---     &    ---     &     ---     &    ---      \\
2002-07-18 & V137 &     ---     &    ---     &    ---       &    ---     &        22187 &   68.2 $\pm$   8.1 &     ---      &    ---     &     ---     &    ---     &     ---     &    ---     &     ---     &    ---      \\
2002-07-18 & V137 &     ---     &    ---     &    ---       &    ---     &        22172 &   78.9 $\pm$   9.0 &     ---      &    ---     &     ---     &    ---     &     ---     &    ---     &     ---     &    ---      \\
2003-09-07 & C1224 &     ---     &    ---     &    ---       &    ---     &        23659 &   90.0 $\pm$   9.0 &     ---      &    ---     &     ---     &    ---     &     ---     &    ---     &     ---     &    ---      \\
2003-11-03 & C1049 &     ---     &    ---     &    ---       &    ---     &        19008 &   96.0 $\pm$   8.0 &     ---      &    ---     &     ---     &    ---     &     ---     &    ---     &     ---     &    ---      \\
2003-11-03 & C1049 &     ---     &    ---     &    ---       &    ---     &        16960 &   97.0 $\pm$   9.4 &     ---      &    ---     &     ---     &    ---     &     ---     &    ---     &     ---     &    ---      \\
2003-11-04 & C1049 &     ---     &    ---     &    ---       &    ---     &        22796 &  109.0 $\pm$  11.0 &     ---      &    ---     &     ---     &    ---     &     ---     &    ---     &     ---     &    ---      \\
2003-11-04 & C1049 &     ---     &    ---     &    ---       &    ---     &        21056 &  107.0 $\pm$  10.0 &     ---      &    ---     &     ---     &    ---     &     ---     &    ---     &     ---     &    ---      \\
2003-11-08 & C1049 &     ---     &    ---     &    ---       &    ---     &     ---       &    ---    &         8256 &   98.2 $\pm$   6.3 &         5056 &   78.0 $\pm$   8.5 &     ---     &    ---     &     ---     &    ---      \\
2004-04-02 & C1424 &     ---     &    ---     &    ---       &    ---     &     ---       &    ---    &         8640 &   78.9 $\pm$   7.2 &         4800 &   71.4 $\pm$   4.7 &         2368 &   52.0 $\pm$   2.2 &         1384 &   52.2 $\pm$   2.0  \\
2004-05-20 & CX065 &     ---     &    ---     &    ---       &    ---     &        22087 &  108.0 $\pm$   9.9 &     ---      &    ---     &     ---     &    ---     &     ---     &    ---     &     ---     &    ---      \\
2004-06-13 & V176 &     ---     &    ---     &    ---       &    ---     &        22209 &   71.7 $\pm$   1.2 &     ---      &    ---     &     ---     &    ---     &     ---     &    ---     &     ---     &    ---      \\
2004-06-13 & V176 &     ---     &    ---     &    ---       &    ---     &        22195 &   67.1 $\pm$   1.1 &     ---      &    ---     &     ---     &    ---     &     ---     &    ---     &     ---     &    ---      \\
2004-08-01 & V176B &     ---     &    ---     &    ---       &    ---     &        22209 &   77.7 $\pm$   9.5 &     ---      &    ---     &     ---     &    ---     &     ---     &    ---     &     ---     &    ---      \\
2004-08-01 & V176B &     ---     &    ---     &    ---       &    ---     &        22195 &   76.5 $\pm$   9.5 &     ---      &    ---     &     ---     &    ---     &     ---     &    ---     &     ---     &    ---      \\
2005-03-14 & C1368 &     ---     &    ---     &    ---       &    ---     &        22193 &  446.0 $\pm$  17.0 &     ---      &    ---     &     ---     &    ---     &     ---     &    ---     &     ---     &    ---      \\
2005-04-20 & C1368 &     ---     &    ---     &    ---       &    ---     &        22194 &  454.0 $\pm$  20.0 &     ---      &    ---     &     ---     &    ---     &     ---     &    ---     &     ---     &    ---      \\
2005-05-25 & C1321 &        88632 &   40.0 $\pm$   7.0 &    ---       &    ---     &     ---       &    ---    &     ---      &    ---     &     ---     &    ---     &     ---     &    ---     &     ---     &    ---      \\
2005-10-27 & C1049 &     ---     &    ---     &    ---       &    ---     &        18752 &   87.7 $\pm$   9.3 &     ---      &    ---     &     ---     &    ---     &     ---     &    ---     &     ---     &    ---      \\
2005-10-27 & C1049 &     ---     &    ---     &    ---       &    ---     &        21056 &   80.6 $\pm$   9.8 &     ---      &    ---     &     ---     &    ---     &     ---     &    ---     &     ---     &    ---      \\
2005-11-13 & C1049 &     ---     &    ---     &    ---       &    ---     &     ---       &    ---    &         8640 &   60.7 $\pm$   3.8 &         4800 &   79.7 $\pm$   2.0 &     ---     &    ---     &     ---     &    ---      \\
2006-04-29 & C1049 &     ---     &    ---     &    ---       &    ---     &        18752 &   99.6 $\pm$   1.1 &     ---      &    ---     &     ---     &    ---     &     ---     &    ---     &     ---     &    ---      \\
2006-04-29 & C1049 &     ---     &    ---     &    ---       &    ---     &        21056 &   88.7 $\pm$   9.4 &     ---      &    ---     &     ---     &    ---     &     ---     &    ---     &     ---     &    ---      \\
2006-06-08 & CX114 &     ---     &    ---     &    ---       &    ---     &        22193 &  448.0 $\pm$  23.9 &     ---      &    ---     &     ---     &    ---     &     ---     &    ---     &     ---     &    ---      \\
2006-06-08 & CX114 &     ---     &    ---     &    ---       &    ---     &        20160 &   57.0 $\pm$  15.0 &     ---      &    ---     &     ---     &    ---     &     ---     &    ---     &     ---     &    ---      \\
2006-06-21 & C1049 &     ---     &    ---     &    ---       &    ---     &     ---       &    ---    &         8640 &   67.3 $\pm$   1.1 &         4800 &   84.6 $\pm$   1.0 &     ---     &    ---     &     ---     &    ---      \\
2006-09-14 & C1341 &     ---     &    ---     &    ---       &    ---     &     ---       &    ---    &     ---      &    ---     &     ---     &    ---     &     ---     &    ---     &         1384 &   35.0 $\pm$   9.2  \\
2006-09-14 & C1341 &     ---     &    ---     &    ---       &    ---     &     ---       &    ---    &     ---      &    ---     &     ---     &    ---     &     ---     &    ---     &         1418 &   56.5 $\pm$   5.8  \\
2007-06-24 & VX013A &     ---     &    ---     &    ---       &    ---     &        22316 &  138.0 $\pm$   7.5 &     ---      &    ---     &     ---     &    ---     &     ---     &    ---     &     ---     &    ---      \\
2007-06-24 & VX013A &     ---     &    ---     &    ---       &    ---     &        22348 &  147.0 $\pm$   7.2 &     ---      &    ---     &     ---     &    ---     &     ---     &    ---     &     ---     &    ---      \\
2008-04-21 & C1803 &     ---     &    ---     &    ---       &    ---     &     ---       &    ---    &         8640 &   56.1 $\pm$   2.8 &         4800 &   65.4 $\pm$   5.8 &         2368 &   57.4 $\pm$   5.3 &         1384 &   64.0 $\pm$   5.2  \\
2008-08-16 & C1750 &     ---     &    ---     &    ---       &    ---     &        22192 &  241.0 $\pm$  14.6 &     ---      &    ---     &     ---     &    ---     &     ---     &    ---     &     ---     &    ---      \\
2008-10-02 & C184 &     ---     &    ---     &        34496 &   78.1 $\pm$   8.2 &     ---       &    ---    &     ---      &    ---     &     ---     &    ---     &     ---     &    ---     &     ---     &    ---      \\
2008-10-12 & C184 &     ---     &    ---     &    ---       &    ---     &     ---       &    ---    &         8640 &   54.0 $\pm$   5.5 &         4786 &   52.2 $\pm$   6.8 &     ---     &    ---     &     ---     &    ---      \\
2008-10-12 & C184 &     ---     &    ---     &    ---       &    ---     &     ---       &    ---    &     ---      &    ---     &     ---     &    ---     &         2496 &  128.0 $\pm$  34.3 &         1376 &  122.0 $\pm$  32.0  \\
2008-12-22 & C184 &     ---     &    ---     &    ---       &    ---     &     ---       &    ---    &         8640 &   46.0 $\pm$   8.3 &         4786 &   74.0 $\pm$   9.0 &         2496 &  120.0 $\pm$  31.7 &         1376 &   83.1 $\pm$  13.2  \\
2009-01-28 & C184 &     ---     &    ---     &    ---       &    ---     &     ---       &    ---    &         8640 &   60.6 $\pm$   7.3 &         4786 &   74.0 $\pm$   8.4 &         2496 &   71.9 $\pm$  14.3 &         1376 &   82.4 $\pm$   2.2  \\
2009-04-13 & CX172 &     ---     &    ---     &    ---       &    ---     &     ---       &    ---    &         8800 &   49.8 $\pm$   5.2 &         5500 &   41.0 $\pm$   5.1 &     ---     &    ---     &     ---     &    ---      \\
2010-01-16 & C1726 &     ---     &    ---     &    ---       &    ---     &     ---       &    ---    &         9000 &   68.2 $\pm$   5.4 &         5500 &   74.8 $\pm$   6.7 &     ---     &    ---     &     ---     &    ---      \\
2010-01-25 & C184 &     ---     &    ---     &    ---       &    ---     &     ---       &    ---    &     ---      &    ---     &     ---     &    ---     &         2450 &   62.8 $\pm$   1.0 &     ---     &    ---      \\
2010-05-05 & V253 &     ---     &    ---     &    ---       &    ---     &        22316 &   51.0 $\pm$   4.9 &     ---      &    ---     &     ---     &    ---     &     ---     &    ---     &     ---     &    ---      \\
2010-05-07 & V253 &     ---     &    ---     &    ---       &    ---     &     ---       &    ---    &         8425 &   90.6 $\pm$   7.0 &     ---     &    ---     &     ---     &    ---     &     ---     &    ---      \\
2011-05-09 & 1.5B &     ---     &    ---     &    ---       &    ---     &     ---       &    ---    &     ---      &    ---     &     ---     &    ---     &         2100 &   63.0 $\pm$   6.5 &         1418 &   76.3 $\pm$   5.2  \\
2011-05-09 & 1.5B &     ---     &    ---     &    ---       &    ---     &     ---       &    ---    &     ---      &    ---     &     ---     &    ---     &         2100 &   65.0 $\pm$   6.3 &         1613 &   90.6 $\pm$  20.4  \\

\enddata
\begin{flushleft}

\tablecomments{
{\it Column 1:} Starting date of observation. 
{\it Column 2:} Observing program ID.
{\it Columns 3, 5, 7, 9, 11, 13, and 15:} Mean frequency of observed band in
units of MHz. 
{\it Columns 4, 6, 7, 10, 12, 14, and 16:} Integrated flux densities in units of mJy as measured by {\sc uvfit} or {\sc imfit} in {\sc miriad}.}
\end{flushleft}

\end{deluxetable}
\end{landscape}


\end{document}